\documentclass[a4paper,11pt]{article}
\pdfoutput=1
\usepackage{jcappub}

\usepackage{amsmath}
\usepackage{bm}
\usepackage{xcolor,color,colortbl}
\usepackage[utf8]{inputenc}
\usepackage{hyperref}
\usepackage{url}
\usepackage{graphicx}
\usepackage{multirow,bigdelim}
\usepackage{amssymb}
\usepackage[amssymb]{SIunits}
\usepackage{multirow}
\usepackage{bbold}
\usepackage{verbatim}
\usepackage{comment}
\usepackage{caption}
\usepackage{physics}
\usepackage{subcaption}

\definecolor{darkgreen}{HTML}{339933}

\usepackage{scalerel,tikz}
\usetikzlibrary{svg.path}
\definecolor{orcidlogocol}{HTML}{A6CE39}
\tikzset{orcidlogo/.pic={
 \fill[orcidlogocol] svg{M256,128c0,70.7-57.3,128-128,128C57.3,256,0,198.7,0,128C0,57.3,57.3,0,128,0C198.7,0,256,57.3,256,128z};
 \fill[white] svg{M86.3,186.2H70.9V79.1h15.4v48.4V186.2z}
 svg{M108.9,79.1h41.6c39.6,0,57,28.3,57,53.6c0,27.5-21.5,53.6-56.8,53.6h-41.8V79.1z M124.3,172.4h24.5c34.9,0,42.9-26.5,42.9-39.7c0-21.5-13.7-39.7-43.7-39.7h-23.7V172.4z}
 svg{M88.7,56.8c0,5.5-4.5,10.1-10.1,10.1c-5.6,0-10.1-4.6-10.1-10.1c0-5.6,4.5-10.1,10.1-10.1C84.2,46.7,88.7,51.3,88.7,56.8z};
}}
\newcommand\orcidicon[1]{\href{https://orcid.org/#1}{\mbox{\scalerel*{
\begin{tikzpicture}[yscale=-1,transform shape]
\pic{orcidlogo};
\end{tikzpicture}
}{|}}}}

\title{An Analytically Tractable Marked Power Spectrum}

\author[a,b,c]{Haruki Ebina~\orcidicon{0000-0002-1080-0955},}
\author[a,b,c]{and Martin White~\orcidicon{0000-0001-9912-5070}}
\affiliation[a]{Department of Physics, University of California, Berkeley, CA 94720, USA}
\affiliation[b]{Berkeley Center for Cosmological Physics, UC Berkeley, CA 94720, USA}
\affiliation[c]{Lawrence Berkeley National Laboratory, One Cyclotron Road, Berkeley, CA 94720, USA}

\emailAdd{ebina@berkeley.edu}
\emailAdd{mwhite@berkeley.edu}

\abstract{
The increasing precision of cosmology data in the modern era is calling for methods to allow the extraction of non-Gaussian information using tools beyond two-point statistics. 
The marked power spectrum has the potential to extract beyond two-point information in a computationally efficient way while using much of the infrastructure already available for the power spectrum. 
In this work we explore the marked power spectrum from an analytical perspective. In particular, we explore a low-order polynomial for the mark that allows us to better control the theoretical uncertainties and we show that with minimal new degrees of freedom the analytical results match measurements from N-body simulations for both the matter field and biased tracers in redshift space. 
Finally, we show that even within the limited forms of mark that we consider, there are degeneracies that can be broken by inclusion of the marked auto-spectrum or the cross-spectrum with the unmarked field.  We discuss future theoretical developments that would enable us to apply this approach to survey data. 
}

\begin{document}
\maketitle
\flushbottom

\section{Introduction}

The study of large-scale structure is one of our key cosmological probes, capable of providing information on galaxy formation, cosmology and fundamental physics \cite{Bernardeau02,Ferraro22,Ivanov22b,Baumann22}.  For previous surveys the majority of the information has come from the two-point function, either in Fourier space (the power spectrum) or configuration space (the correlation function), see e.g.\ refs.~\cite{Alam21,DESI24-VI} for recent examples.  The near-Gaussianity of the density field on large scales makes this an efficient compression of information, however it is not entirely lossless. With current and next generation of surveys returning measurements of steadily larger volumes and increased precision, it is important to revisit the question of how best to capture non-Gaussian information with tools beyond the simple two-point function of density fields. 

There are multiple avenues towards obtaining non-Gaussian information from the data, the most straight-forward being the inclusion of higher moments such as the bispectrum \cite{Peebles75,Fry82} and trispectrum \cite{Fry78,Hu01}. The bispectrum, in particular, has been used in the most recent generation of surveys \cite{Gil-Marin16,Philcox22b,Ivanov23,DAmico24} returning modest improvements in parameter constraints over the power spectrum due to degeneracy breaking \cite{Bernardeau97,Pires12,Hahn21}. 
While there have been advances in the past few years, computational complexity remains one of the major challenges of using higher-order statistics \cite{Philcox22c,Giri23,Choustikov23,Hou23,Philcox24,Harscouet24}. In addition to numerical difficulties, the requirements on covariances, window functions, sensitivity to systematics or artifacts from the survey operations all need to be carefully evaluated.

Beyond the simple $n$-point functions there are many alternative summary statistics that have been investigated. These include, but are not limited to, clipped density fields \cite{Simpson11}, density-split statistics \cite{Gruen16,Friedrich18,Paillas21,Paillas24}, $k$-nearest neighbors \cite{Banerjee21}, wavelet scattering transforms \cite{Cheng20,Valogiannis22,Eickenberg22,Cheng24}, and modeling of log-normal fields \cite{Rubira21}. Many of these methods have been compared in the recent ``Beyond 2-pt Challenge''  \cite{Krause24}.
While often more computationally efficient than using higher-point statistics, these techniques often suffer from a lack of a well-controlled theoretical model, making the estimation of the total error challenging.  

Marked power spectra, which compute the power spectrum of an overdensity field weighted\footnote{Marked spectra are similar to skew-spectra \cite{Schmittfull15a, Chen24, Hou24} so we shall not consider these separately.} by a ``mark'' (in our case a function of the smoothed overdensity field), offer a computationally inexpensive tool for including additional information to break degeneracies while providing a theoretically well-constrained way to include higher-order information.  Since the marked power-spectrum contains three-or-more powers of the galaxy density field, it directly encodes information of higher-order statistics. One of the great advantages of marked spectra is that they can make use of the survey infrastructure developed for the computation of the power spectrum itself, bypassing the need for many of the validation requirements of other methods. In addition, extensive infrastructure exists to handle systematics in two-point functions. By taking advantage of this we believe that the complexity of survey systematics treatment can be reduced. The formalism was initially introduced by ref.~\cite{Stoyan84,Beisbart02,Skibba06,White09b} and reintroduced by ref.~\cite{White16} which highlighted a particular mark that emphasizes the low-density regions useful for modified gravity studies. This choice of mark has since been applied to data to test for modified gravity \cite{Armijo18, Hernandez-Aguayo18,Satpathy19, Armijo24}. Recently, ref.~\cite{Karcher24} has extended these works by considering several different forms of marks for modified gravity and implementing them against N-body simulations. 

Ref.~\cite{White16} studied the marked correlation function within the context of the Zeldovich approximation, and this was extended in refs~\cite{Philcox20,Philcox21} to one-loop in (Eulerian) perturbation theory.  While the formalism of \cite{Philcox21} used a general power-series expansion of the mark, they focused their attention on the mark used in ref.~\cite{White16}. 
With this mark these authors identified some challenges. On the theoretical front, they have identified non-negligible, large-scale corrections to the linear theory that come from higher-loop-order contributions. Practically, they have also struggled to find consistency between analytical calculations and simulations. 

In this work we wish to revisit the use of marked power spectra in cosmology.  Rather than being focused on testing modified gravity, our interest will instead be to see whether marked spectra can offer a practical means of breaking degeneracies among parameters in our theoretical models while being theoretically well controlled and making the most use of the existing survey infrastructure.  To this end we will consider a different functional form for the mark, in particular an at-most-second-degree polynomial of the smoothed overdensity field. 
The choice of the mark function will allow us to naturally extend previous studies to mediate previous theoretical issues. Our work is similar in spirit, though complementary, to ref.~\cite{Cowell24}.  Those authors explored a general form for the marks that are functions of the smoothed density field by using Gaussian processes, using N-body simulations as the theoretical predictions.  Our focus will be on marks that are well-controlled within perturbation theory, and we will present an analytic model for the predictions.

Marks that are low-order in the smoothed overdensity field, with smoothing radii that are larger than the non-linear scale, are most amenable to an analytic treatment.  We will show that marks extending to high-order in the smoothed overdensity field are more difficult to model and are more susceptible to higher order corrections that can appear even on large-scales, spoiling our ability to reliably model the data with only a few parameters.  Intuitively this behavior can be understood as the product of smoothed overdensity fields probing to more non-linear scales through a chain of (smoothed) convolutions or as higher powers upweighting the tails of the overdensity distribution.

The structure of the paper is as follows. First, in \S\ref{methods} we review the theoretical structure of the marked power-spectrum, leaning heavily on \cite{Philcox20,Philcox21}. In \S\ref{low-k} we follow-up on the theoretical predictions of large-scale corrections to the one-loop theory. In \S\ref{cross} we explore the cross-spectrum between two marked overdensity fields. We confirm our theoretical robustness against mock catalogs built from N-body simulations in \S\ref{mocks}. In \S\ref{degeneracy} we will confirm the existence and viability of degeneracy breaking using the marked power-spectrum. Finally we will conclude with some remarks in \S\ref{conclusion}.  For the purposes of making figures and comparing to simulations we use a particular $\Lambda$CDM cosmology given in detail in \S\ref{mocks}.

\section{Methodology}
\label{methods}

\subsection{The marked spectrum}

\begin{figure}
    \centering
    \includegraphics[width=1\linewidth]{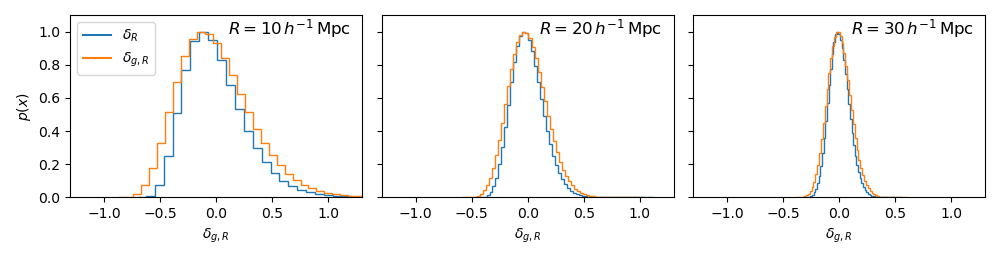}
    \caption{The normalized distribution of the smoothed overdensity, $\delta_{g,R}$ for the matter (blue) and biased tracer (orange) for the Abacus simulations used in this work.
    }
    \label{fig:deltaR}
\end{figure}

The marked power-spectrum is a generalization of the power-spectrum, where one uses the marked density field
\begin{equation}
    \label{rho_M}
    \rho_M(\bm{x}) = m(\bm{x})\rho_g(\bm{x})
\end{equation}
with weights $m$ (called the mark), instead of the galaxy density field $\rho_g$ \cite{White16,Philcox21}. By making the mark a function of the smoothed galaxy overdensity field, the over- and under-dense regions can be weighed differently thus bringing in more information than the 2-point function alone.  Depending upon the mark (see later) this can be done in a theoretically sound, perturbative manner.  In fact if we ensure that the smoothing scale, $R$, is large enough then $\delta_{g,R}$ has no contributions from high $k$ modes.  Much past work on marked statistics for cosmology \cite{White16,Massara21,Philcox20,Philcox21} focused on a mark of the form 
\begin{equation}
    m = \left(\frac{1+\delta_s}{1+\delta_s+\delta_{g,R}}\right)^p
    \label{Cn}
\end{equation}
where $\delta_s$ and $p$ are adjustable parameters, and $\delta_{g,R}(\bm{k})=W_R(k)\delta_g(\bm{k})$ is the smoothed galaxy overdensity field with the smoothing window $W_R$. For this paper, we choose $W_R$ to be a Gaussian window, while some works have explored the use of a top-hat window \cite{Massara21,Philcox20}. The (normalized) distribution of $\delta_{g,R}$ for the smoothing radii $R=\{10,20,30\}\,h^{-1}\,{\rm Mpc}$ that we consider in this work are shown in Figure \ref{fig:deltaR} for both matter and the biased tracer used in this work. With positive $p$ Eq.~(\ref{Cn}) emphasizes under-dense regions where screening mechanisms in some modified gravity models predict observational signatures. However, the formalism of marked statistics is not limited to this mark as the mark can be any function of the smoothed density field. In particular, it is useful to express the mark in a Taylor expansion\footnote{Note that this differs from ref.~\cite{Philcox21} by a factor of $(-1)^n$, which we have included in $C_n$.} \cite{White16,Philcox21}
\begin{equation}
    m = \sum_n C_n \delta_{g,R}^n(\bm{x}) \qquad .
\end{equation}
with $C_n$ the expansion coefficients that we specify. 
For this work we limit ourselves to marks that are quadratic in $\delta_{g,R}$, for reasons that will become apparent shortly.  Also, for the purposes of this paper we shall assume that the fiducial cosmology assumed to convert angles and redshifts to distances is the ``correct'' one and thus ignore ``the Alcock-Paczynski (A-P) effect''.  In any application to real data such terms need to be included in our theoretical model when defining the smoothed overdensity field.  Our purpose here is to first investigate the utility of this approach -- a proper treatment of A-P effects would be necessary before application of marked power-spectra to data.

\subsection{The one-loop marked spectrum}

From Eqn. \ref{rho_M} the definition for the marked overdensity field follows as
\begin{equation}
    \delta_M = \frac{\rho_M}{\expval{\rho_M}}-1 = \frac{m}{\Bar{m}}(1+\delta_g)-1
\end{equation}
where $\bar{m}=\expval{\rho_M}/\expval{\rho_g}$ can be measured from data or simulations. 
Within perturbation theory we can expand $\delta_M$ as \cite{Philcox21}
\begin{align}
    \Bar{m}[1+\delta_M(\bm{x})] &= \sum_n C_n \delta_{g,R}^n(\bm{x})[1+\delta_g(\bm{x})] \\
    &= \sum_n C_n \left[ \sum_{m=0}^\infty \delta_{g,R}^{(m)}(\bm{x}) \right]^n \left[ 1 + \sum_{j=1}^\infty \delta_g^{(j)}(\bm{x}) \right] \\
    &\equiv \left[ C_0+\sum_{n=1}^\infty \delta_M^{(n)}(\bm{x}) \right]
\end{align}
where $\delta_M^{(n)}$ is the contribution from order $n$, and in our convention contains a factor of $\bar{m}$. This indicates that each $\delta_M^{(n)}$ consist of sums of $\delta_g$, $\delta_{g,R}$, and their products. For example, 
\begin{equation}
    \delta_M^{(2)}(\bm{x}) = \left[ C_0\delta_g^{(2)} + C_1 \delta_{g,R}^{(2)} + C_1 \delta_{g,R}^{(1)}\delta_{g}^{(1)} + C_2 \delta_{g,R}^{(1)}\delta_{g,R}^{(1)}\right](\bm{x})
\end{equation}
We shall assume that $R$ is chosen sufficiently large that higher powers in the expansion are suppressed.

Making the standard Einstein-de Sitter approximation for the PT kernels\footnote{Throughout we shall adopt the notation $\int_{\mathbf{k}_1\cdots\mathbf{k}_n}=\int\prod_{i=1}^n d^3k_i/(2\pi)^3$.}
\begin{equation}
  \delta_g^{(n)}(\mathbf{k}) = \int_{\mathbf{k}_1\cdots\mathbf{k}_n}
  (2\pi)^3\delta_D\left(\sum_i\mathbf{k}_i-\mathbf{k}\right)
  Z_n(\mathbf{k}_1,\cdots,\mathbf{k}_n;\mathbf{k})
  \delta^{(1)}(\mathbf{k}_1)\cdots\delta^{(1)}(\mathbf{k}_n)
\end{equation}
with $Z_n$ the standard kernels of Eulerian perturbation theory \cite{Bernardeau02}, for example 
\begin{align}
    Z_1(\mathbf{k}_1) &= b_1 + f\mu_1^2 \\
    Z_2(\mathbf{k}_1,\mathbf{k}_2) &= b_1 F_2(\mathbf{k}_1,\mathbf{k}_2) + f\mu_k^2G_2(\mathbf{k}_1,\mathbf{k}_2) + \frac{b_2}{2} + b_s\left(\frac{(\mathbf{k}_1\cdot\mathbf{k}_2)^2}{k_1^2k_2^2}-\frac{1}{3} \right) \\ 
    &+ \frac{fk\mu_k}{2}\left[\frac{\mu_1}{k_1}(b_1+f\mu_2^2) + \frac{\mu_2}{k_2}(b_1 + f\mu_1^2) \right]
\end{align}
The smoothed density field is then simply $W_R(k)\delta_g$ and can be similarly expanded perturbatively.

\begin{figure}
\begin{subfigure}{.2\textwidth}
  \centering
  \includegraphics[width=\linewidth]{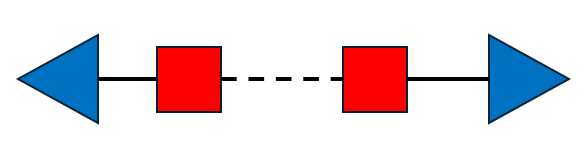}
  \caption{$M_{11}$}
\end{subfigure}
\begin{subfigure}{.395\textwidth}
  \centering
  \includegraphics[width=\linewidth]{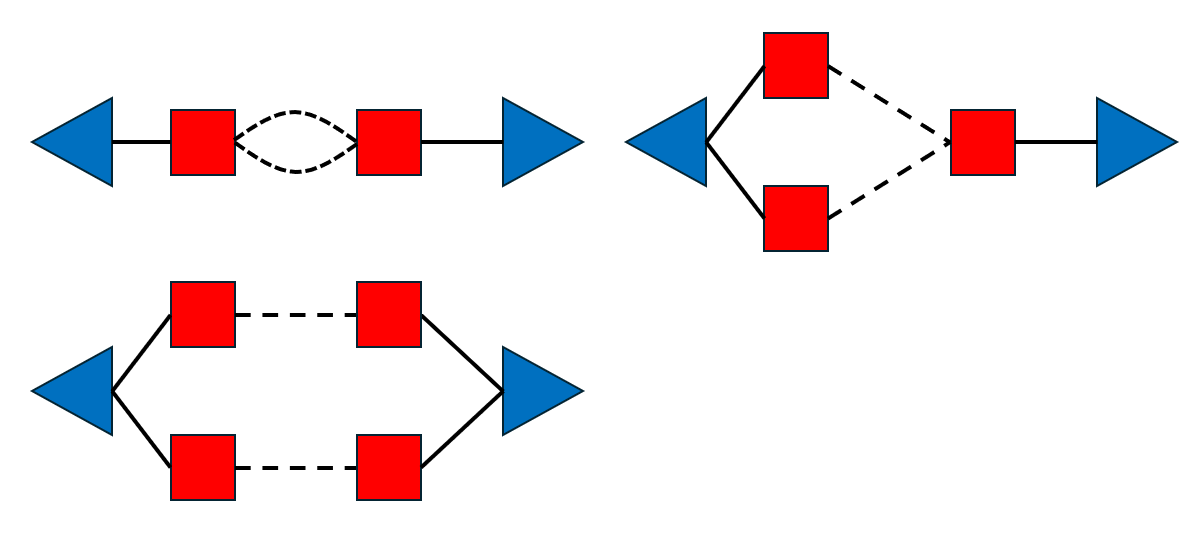}
  \caption{$M_{22}$}
\end{subfigure}
\begin{subfigure}{.395\textwidth}
  \centering
  \includegraphics[width=\linewidth]{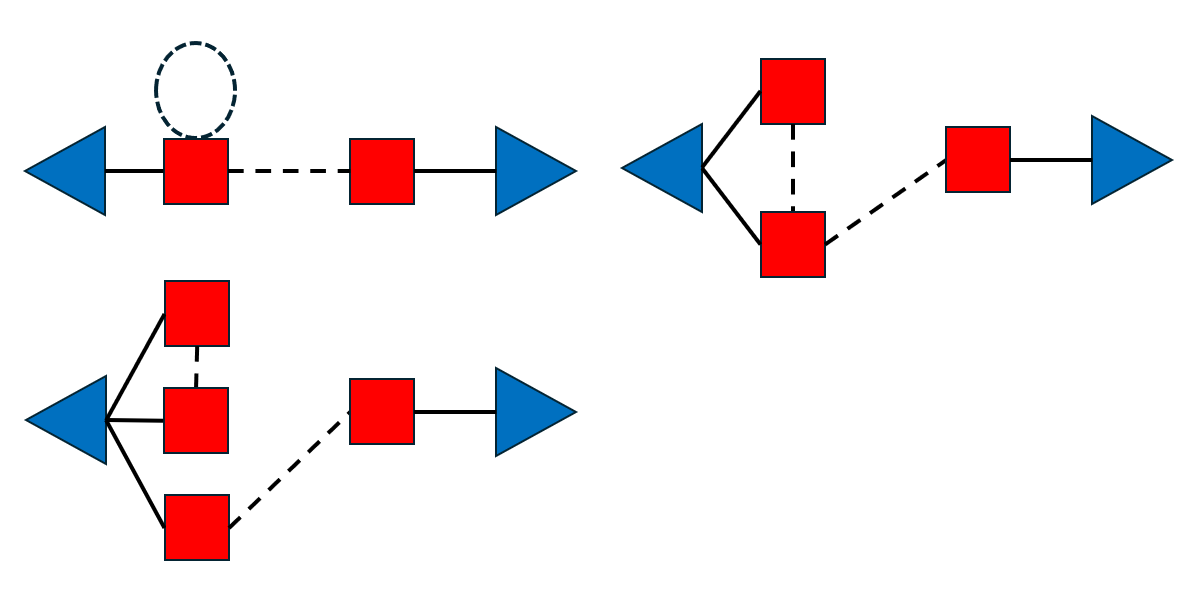}
  \caption{$M_{13}$}
\end{subfigure}
\caption{Diagrams of contractions contributing to $M(\bm{k})$. Each diagram consists of a contraction between two $\delta_M$ operators, represented by the blue triangles. Each $\delta_M$ operator consists of $\delta_g$ operators, represented by the red squares stemming from the blue triangles. Finally, each diagram is characterized by the contractions between $\delta_g$ operators, represented by the dashed lines connecting the red squares. Thus, diagrams with two, three, or four red squares ($\delta_g$ operators) each contribute to $M^A$, $M^B$, or $M^C$. Note that the diagrams do not distinguish between $\delta_g$ and $\delta_{g,R}$. Each panel of the figure represents, respectively, the diagrams contributing to (a) $M_{11}$, (b) $M_{22}$, and (c) $M_{13}$. }
\label{fig:diagrams}
\end{figure}

With the marked overdensity field written perturbatively, the marked power-spectrum can be also be expanded perturbatively as a sum of components similar to the power-spectrum
\begin{equation}
    M_{11} (\bm{k}) = \expval{\delta_M^{(1)}\delta_M^{(1)}} \qquad M_{13} (\bm{k}) = \expval{\delta_M^{(1)}\delta_M^{(3)}} \qquad M_{22} (\bm{k}) = \expval{\delta_M^{(2)}\delta_M^{(2)}}
\end{equation}
\begin{equation}
    \bar{m}^2 M = M_{11} + 2 M_{13} + M_{22}
\end{equation}
where we note that $\delta_M$ and hence $M_{ij}$ contain powers of $\bar{m}$ while we have defined $M$ to factor out $\bar{m}$.
It is natural to decompose the contributions into terms that include contractions between two, three, or four $\delta_g^{(n)}$ operators. A diagrammatic description captures this naturally in Figure \ref{fig:diagrams}. Notice that each diagram consists of two, three, or four $\delta_g$ operators (red squares), indicating the natural categorization of terms. Let us label the two, three and four $\delta_g$ contributions as $M^A$, $M^B$, and $M^C$ respectively \cite{Philcox21}. 

Going forward, it is convenient to define the following expressions, which shall recur frequently,
\begin{align}
    C_{\delta_M}(k_1) &= C_0+C_1 W_R(k_1) \\
    C_{\delta_M^2}(k_1,k_2) &= C_2 W_R(k_1)W_R(k_2) + \frac{C_1}{2}\left[ W_R(k_1)+W_R(k_2) \right] \\
    C_{\delta_M^3}(k_1,k_2,k_3) &= \frac{C_2}{3}\left[ W_R(k_1)W_R(k_2) + W_R(k_1)W_R(k_3) + W_R(k_2)W_R(k_3) \right] \nonumber \\
    &+ C_3 W_R(k_1)W_R(k_2)W_R(k_3)
\end{align}
to encapsulate the possible combinations of smoothed and unsmoothed density fields.

The terms with two $\delta_g$ operators are those manifest in the unmarked power-spectrum, $P(k,\mu)$, and thus contribute a term in $M$ proportional to the power-spectrum
\begin{align}
\begin{split}
    M_{11} &= M_{11}^A = C_{\delta_M}^2(k)Z_1(\bm{k})^2 P_L(k) = C_{\delta_M}^2(k) P_{11}(\bm{k})\\
    M_{13} &\supset M_{13}^A = 3 C_{\delta_M}^2(k)Z_1(\bm{k}) \int_{\bm{p}} Z_3(\bm{p},-\bm{p},\bm{k})P_L(p) = C_{\delta_M}^2(k) P_{13}(\bm{k}) \\
    M_{22} &\supset M_{22}^A = 2 C_{\delta_M}^2(k) \int_{\bm{p}} Z_2(\bm{p},\bm{k}-\bm{p})^2 P_L(p) P_L(|\bm{k}-\bm{p}|) = C_{\delta_M}^2(k) P_{22}(\bm{k}) 
\end{split}
\label{M_A}
\end{align}
where we have labelled all contributions with $A$ according to the above categorization. 

Those with three $\delta_g$ operators are terms that contribute a term proportional to the tree-level bispectrum, with one of the momenta integrated over due to the nature of observing a two-point function instead of a three-point function. These give the contributions
\begin{align}
\begin{split}
    M_{13} &\supset M_{13}^B = 4C_{\delta_M}(k)Z_1(\bm{k})P_L(k) \int_{\bm{p}} C_{\delta_M^2}(p,|\bm{k}-\bm{p}|) Z_1(\bm{p})Z_2(\bm{k},-\bm{p})P_L(p)   \\
    M_{22} &\supset M_{22}^B = 4C_{\delta_M}(k)\int_{\bm{p}} C_{\delta_M^2}(p,|\bm{k}-\bm{p}|) Z_2(\bm{p},\bm{k}-\bm{p})Z_1(\bm{p})Z_1(\bm{k}-\bm{p})P_L(p)P_L(|\bm{k}-\bm{p}|)    
\end{split}
    \label{M_B}
\end{align}
The final set of contribution including four $\delta_g$ operators are terms that contribute convolutions of the linear theory power-spectrum. 
\begin{align}
\begin{split}
    M_{13} &\supset M_{13}^C = 3 C_{\delta_M}(k) Z_1(\bm{k})^2 P_L(k) \int_{\bm{p}}C_{\delta_M^3}(p,p,k)Z_1(\bm{p})Z_1(-\bm{p}) P_L(p)  \\
    M_{22} &\supset M_{22}^C = 2\int_{\bm{p}} \left[C_{\delta_M^2}(p,|\bm{k}-\bm{p}|) Z_1(\bm{p})Z_1(\bm{k}-\bm{p})\right]^2 P_L(p)P_L(|\bm{k}-\bm{p}|) 
\end{split}
    \label{M_C}
\end{align}
We refer the reader to ref.~\cite{Philcox21} for further details\footnote{The final $M_{13}^B$ expression of ref.~\cite{Philcox21} in Appendix A has a factor of two error .}.

\subsection{Small-scale behavior} \label{small-scale}

A significant difference between the marked and unmarked spectra is that the mark introduces zero-lag correlators between $\delta_g$ and $\delta_{g,R}$, i.e.\ fields that are evaluated at the same point in space. An example of this is the ``three-point'' contribution $M_{13}^B$, including terms such as $\expval{\delta_g^{(1)}(\bm{x_1})\delta_{g,R}^{(2)}(\bm{x_1})\delta_g^{(1)}(\bm{x_2})}$ that do not appear in the usual $P(\bm{k})$.

These ``higher-point'' contributions are explicitly integrations over higher-point functions, which motivates the statement that the marked power-spectrum can extract beyond-two-point information within the framework of two-point correlators.  The presence of more zero-lag correlators than in the power-spectrum may appear concerning, since they could depend upon small-scale physics that is not well understood in a complex way. However, since the marked overdensity field can only have one power of the unsmoothed density field, and the smoothed density field contains no contributions from $k>R^{-1}$, there will be no (new) issues due to UV divergence. Specifically, contractions between $\delta_g$ and $\delta_{g,R}$, or between two $\delta_{g,R}$'s will not lead to UV sensitivity that is not already present in the power spectrum, viz.~the self-contractions (e.g.~$Z_2(\bm{p},-\bm{p})$), which introduce counterterms already in the power-spectrum \cite{Philcox20}. To one-loop order this means that the counter-term contribution is identical, up to overall smoothing $C_{\delta_M}^2(k)$, to the rest of the power-spectrum contributions to the marked power-spectrum. Specifically, this is
\begin{equation}
    \bar{m}^2M(k,\mu) \supset C_{\delta_M}^2(k)\ \alpha_{2n} k^2 \mu^{2n} P_L
\end{equation}
As there are no new counterterm contributions to the marked power-spectrum, the nuisance terms $\alpha_\ell$ are shared between the unmarked and marked-spectra, as well as between marked-spectra.

Unlike the counterterms that handle the UV-sensitivity, the stochastic contributions need not be shared between the marked and unmarked power spectrum, or between marked power spectra. In fact, it is known that products of $n$ operators at the same point necessarily involve careful treatment of field stochasticity. As with the counterterms, we start with the stochastic terms present within the power spectrum
\begin{equation}
    \bar{m}^2M(k,\mu) \supset C_{\delta_M}^2(k) P_{st.} = C_{\delta_M}^2(k) (N + N_2 k^2 \mu^2)
    \label{M_A_correction}
\end{equation}
where $P_{st.}$ is the stochastic contribution to the power spectrum, which includes the shot-noise contribution and first-order Finger of God correction (see e.g.~discussion in ref.~\cite{Maus24b}). 
In addition to this, we must consider stochastic corrections to the `three-point' (Eqn.~\ref{M_B}) and `four-point' (Eqn.~\ref{M_C}) contributions to the marked power spectrum.  These are listed in detail in Appendix \ref{app:stochastic} and arise from small scale contributions to the overdensity field like
\begin{equation}
    \delta^{\rm short} \supset d_1\, \epsilon  + d_1\, \epsilon f\mu^2 \theta + d_2\, b_1\epsilon \delta 
    \label{eqn:delta_short}
\end{equation}
where $d_1$ and $d_2$ are unknown coefficients and $\epsilon$ is a stochastic field with $\langle\epsilon\delta\rangle=0$. 
This gives rise to the contributions from $\expval{ (\delta) (d_1 \epsilon)(d_2 b_1 \epsilon \delta)}=d_1 d_2 b_1\expval{\epsilon^2}\expval{\delta^2}$, $\expval{ (\delta) (d_1 \epsilon)(d_1 f \mu^2\epsilon \theta)}=d_1^2 f\mu^2 \expval{\epsilon^2}\expval{\delta^2}$ and $\expval{(d_1 \epsilon)^3}$.  These additional terms, and their 4-point analogues, add contributions to Eqs.~(\ref{M_B}, \ref{M_C}) as listed in Appendix \ref{app:stochastic}.

In addition to this tree-level result, some authors include corrections that are formally one-loop to handle Fingers of God (FoG) \cite{Ivanov19} in the bispectrum. To lowest order this involves the substitution
\begin{equation}
    Z_1(\bm{k}) \to Z_1^{\rm FoG}(\bm{k}) \equiv b_1 + f\mu^2 + c_1 k^2 \mu^2 + c_2 k^2 \mu^4
\end{equation}
where $c_n$ are defined by $c_n k^2 \mu^{2n}\delta \subset \delta^{\rm short}$ and are constrained by the power-spectrum counterterms $\alpha_{2n}$. The $c_2$ term above is frequently dropped due to a degeneracy with $c_1$ for bispectrum monopoles. 
Whilst other one-loop corrections (such as the counterterm correction $c_0 k^2$) are also allowed, these terms are often neglected.  Such terms would also contribute to Eqs.~(\ref{M_B}, \ref{M_C}) and yield an array of new terms with a complex mixture of contributions including, for example, 
\begin{align}
\begin{split}
    M_{13}^B &\supset 2C_{\delta_M}(k)B_{\rm shot} b_1 (c_1 k^2 \mu^2 + c_2 k^2 \mu^4)  P_L(k)  \int_{\bm{p}} C_1W_R(p) \propto C_{\delta_M}(k)k^2 P_L(k) \\
    M_{13}^B &\supset 2C_{\delta_M}(k)B_{\rm shot} b_1 \int_{\bm{p}} C_1W_R(p)(c_1 p^2 \mu_p^2 + c_2 p^2 \mu_p^4) P_L(p)\propto  C_{\delta_M}(k)
\end{split}
\label{eqn:1-loop-stochastic}
\end{align}
For the `four-point' contributions, one must perform the substitution $Z_1\to Z_1^{\rm FoG}$ similar to the bispectrum. 
Finally there are new small-scale corrections to $\delta$ that are allowed by symmetries, such as $k^2 \epsilon \delta$. 

In practice, we expect many of these terms to be nearly degenerate on large scales. In the absence of smoothed fields, symmetry arguments alone will restrict corrections to even powers of $k$ and $\mu$ either alone or combined with components of the power spectrum. Although in the marked theory we have already seen that the smoothing kernels enter into this nontrivially and allow for complex scale and angular dependencies, we still expect many of these corrections to be degenerate for the limited scale and angular dependence that we probe. In practice, we find that a simple ansatz of the form 
\begin{equation}
    \bar{m}^2M(k,\mu) \supset C_{\delta_M}(k) k^2 (N_{m,2}^{(0)} + \mu^2 N_{m,2}^{(2)} )
    \label{ansatz}
\end{equation}
with $N_{m,2}^{(0)}$ and $N_{m,2}^{(2)}$ both new degrees of freedom, retains the key features of the above array of corrections and is sufficient to demonstrate agreement between analytical results and simulations. In total, the stochastic corrections to the marked power spectrum are the tree-level corrections in Eqns.~\ref{M_A_correction}, \ref{M_B_correction}, and \ref{M_C_correction}, and the ansatz of Eqn.~\ref{ansatz}. 

Thus in total, the marked theory has the typical set of nuisance parameters and four stochastic degrees of freedom, $\{B_0, B_{\rm shot}, N_{m,2}^{(0)}, N_{m,2}^{(2)} \}$, with the latter two an ansatz for the degenerate combination of several terms. In anticipation of using the marked spectrum with the power spectrum in analyses of observational data, when comparing against simulations in what follows we will fit the shared nuisance parameters using the power spectrum, thus varying only the stochastic parameters for the marked power spectrum\footnote{When the smoothing scales are small and tracer density is sparse, the predictions need to be corrected for discreteness effects. We will refer the reader to Section 6 of ref.~\cite{Karcher24} for a detailed discussion.}.

\subsection{Long wavelength displacements}

Finally, we will leave the full treatment of IR-resummation in the marked power-spectrum for future work. Instead, we adopt the same approximation as in refs.~\cite{Philcox20,Philcox21} including IR-resummation only for each contribution of the linear and non-linear power spectra present by performing the substitutions
\begin{align}
    P_L(k) &\to  P_L^{nw} + e^{-\frac{1}{2}k^2 \Sigma(\mu)^2} P_L^w  \\
    P_{NL}(k,\mu) &\to  P_L^{nw} + P_{\rm loop}^{nw} + e^{-\frac{1}{2}k^2 \Sigma(\mu)^2} \left( (1+\frac{1}{2}k^2 \Sigma(\mu)^2) P_L^w + P_{\rm loop}^w \right)
\end{align}
where $P_{\rm loop}$ represents the power-spectrum contribution arising from one-loop correction.
Similar to the treatment of stochastic terms, we will see that there is satisfactory agreement without a complete treatment thus making the development of a full model of IR effects less pressing.

\section{Low-k correction} \label{low-k}

While theoretically stable, the marked power spectrum encounters a challenge in practice. This is due to perturbatively higher order terms entering even at low-$k$ \cite{Philcox21}. In other words, the predictability at low-$k$ providing the constraining power for linear theory parameters can be jeopardized. The marked power spectrum is not the only instance of such low-$k$ corrections appearing in theory. In general, this can happen when taking the compressed statistics of a field that is beyond-linear with respect to the overdensity field. An well-known example of this is the Lyman-$\alpha$ flux two-point function, which uses the flux field $F(\bm{x})=e^{-\tau(\bm{x})}$ where optical depth $\tau$ is dealt with like $\delta_g$ by using bias expansions. Here, a similar correction term to the Kaiser power spectrum has been identified \cite{McDonald00,Mcdonald03,Seljak12,Cieplak16,Chen21,Ivanov24}; indeed this is the main reason why observations of the Ly$\alpha$ forest fit for both a ``flux bias'' and a ``velocity bias'' \cite{DESI24-IV}.
Similar concerns apply to voids, clipped statistics, etc.\ \cite{Seljak12,Chuang17}.  A major advantage of the marked statistics is that we limit the number of unsmoothed fields, controlling UV-unsafe corrections leading to a controlled modification to linear theory.  This will be confirmed in practice against simulations in \S\ref{mocks}. 

These low-$k$ corrections enter as a result of multiple fields in contact, e.g.~$\delta_g(\bm{x_1})\delta_{g,R}(\bm{x_1})$, which introduce UV-safe contact terms that are perturbatively well-behaved but do not scale as $k^{2n}$. 
Ref.~\cite{Philcox21} has identified two categories of such low-$k$ contributions.

\begin{figure}
\begin{subfigure}{.395\textwidth}
  \centering
  \includegraphics[width=\linewidth]{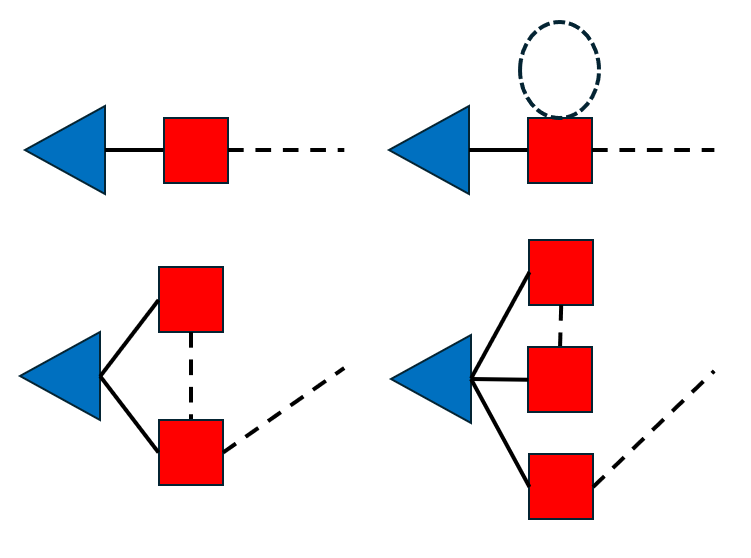}
  \caption{One-loop}
  \label{fig:1leg1}
\end{subfigure}
\begin{subfigure}{.595\textwidth}
  \centering
  \includegraphics[width=\linewidth]{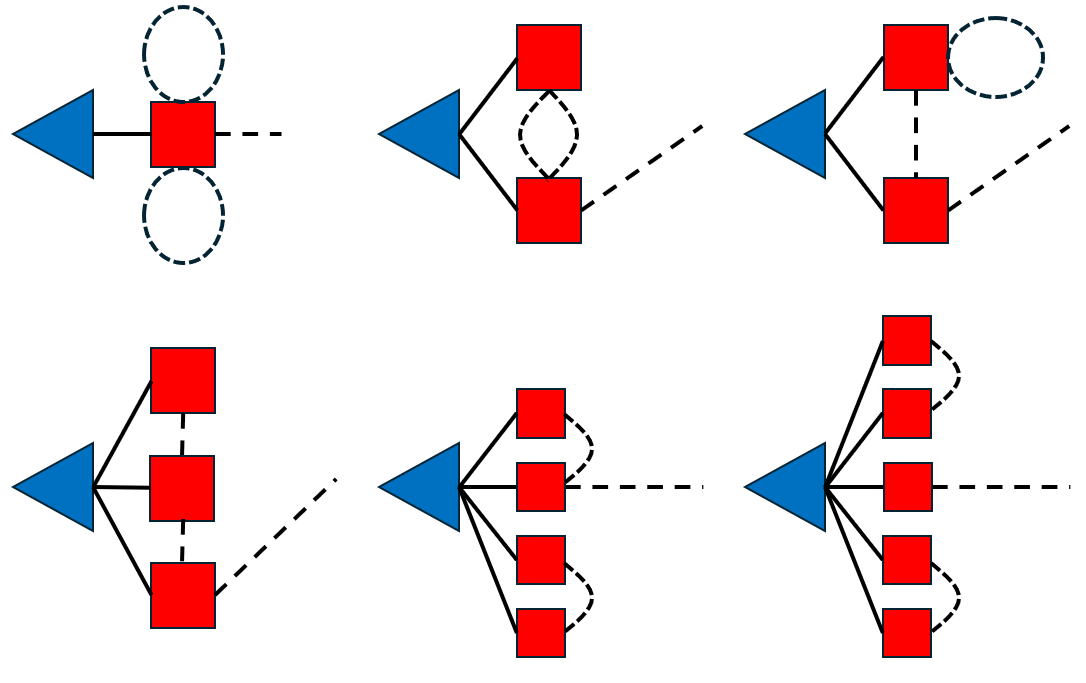}
  \caption{Beyond one-loop}
  \label{fig:1leg2}
\end{subfigure}
\caption{Diagrams of $\delta_M^{(n)}$ with one external `leg', that contribute terms similar to the linear theory power spectrum to the marked power spectrum. The left and right panels each show contributions within and beyond one-loop, respectively. The beyond one-loop diagrams are non-exhaustive, even at two-loop. Note that by constraining the order of the mark, one can greatly reduce the number of such contributions beyond one-loop.}
\label{fig:1leg}
\end{figure}

The first category consists of contributions that are proportional to $P_L$ at low-$k$. These can arise from contributions such as $M_{13}^C\supset\expval{\delta^{(1)}_{g,R,1} \delta^{(1)}_{g,R,1} \delta^{(1)}_{g,1} \delta^{(1)}_{g,2}}$ (Eqn.~\ref{M_C}) that contract $\delta_M$'s with one external `leg', i.e.~one uncontracted $\delta_0$. As seen in Figures \ref{fig:diagrams} and \ref{fig:1leg}, to one-loop order these are $\delta_M^{(1)}$ and $\delta_M^{(3)}$ operators, included in the $M_{11}$ and $M_{13}$ diagrams, but at higher-loop-order the potential contributions combinatorically increase for any odd-order $\delta_M^{(2n+1)}$. These terms end up including additional biases to the monopole and quadruple components of the linear theory power spectrum, directly endangering constraints on large-scale bias, $b$, or growth rate, $f$. 

The second set of contributions are those that go as a constant at low-k. This arise from contractions such as $M_{22}^B\supset\expval{\delta^{(2)}_{g,1}\delta^{(1)}_{g,2}\delta^{(1)}_{g,R,2}}$ that contract $\delta_M$'s with two external `legs'. As seen in Figure \ref{fig:diagrams}, to one-loop order these arise from $\delta_M^{(2)}$. Although these are theoretically of equal standing to the first category of contributions, they are degenerate with stochastic terms and thus of less concern to us, practically. When involving multiple marked power-spectra (including a combination of unmarked and marked spectra), there is a possibility that the stochastic terms differ non-negligibly. While we investigate this in \S\ref{mocks}, a detailed study of new stochastic terms will be left for a future study. 

As mentioned above, these low-$k$ corrections are perturbative and enter as a series in the variances of the smoothed field (e.g.~$\sigma_{RR}^2 = \int_p W_R^2(p) P_L(p)$) or covariances of between the unsmoothed and smoothed field (e.g.~$\sigma_{R}^2 = \int_p W_R(p) P_L(p)$). Importantly note that the UV-unsafe variance of the unsmoothed field (e.g.~$\sigma^2 = \int_p P_L(p)$) does not appear, as there is only one unsmoothed field at each point.  If we choose the smoothing large enough, then these variances are small and thus the deviations from the normal low-$k$ limit are small and of theoretically understood amplitude.

To get the first category of contributions, scaling as $P_L$ at low-$k$, we need $\delta_M$ contributions with one external leg. In an arbitrary mark, these contributions can be a combination of $\delta_g$ and unrestricted copies of $\delta_{g,R}$. To one-loop order the mark's dependence on powers of $\delta_{g,R}$ beyond the 4th cannot be analytically included, making application of the theory harder. Simplifying the mark, e.g.\ to linear order, greatly reduces the number of potential combinations. As self-contractions of operators (e.g.~$Z_n(\bm{p},-\bm{p},\dots)$) lead to counterterms irrelevant at large scales (hence absent from $M(\bm{k})$ at low-$k$, or yielding small corrections due to the smoothing kernel), we consider mostly contributions of the type $\delta_g^{(n)}\delta_{g,R}^{(n\pm 1)}\subset \delta_M^{(2n\pm 1)}$ that can be written out as 
\begin{multline}
    \delta_g^{(n)}\delta_{g,R}^{(n + 1)} = \int_{\bm{k}_1\cdots\bm{k}_n} Z_n(-\bm{k}_1,\dots,-\bm{k}_n)Z_{n+1}(\bm{k}_1,\dots,\bm{k}_n,\bm{k})W_R\left(\sum_{i=1}^n \bm{k}_i+\bm{k}\right) \\
    P_L(k_1)\dots P_L(k_n) \delta_L(\bm{k})
\end{multline}
and similarly for $\delta_g^{(n)}\delta_{g,R}^{(n-1)}$. Since the smoothing kernel over a sum of vectors necessarily contain smoothing of each vector, 
\begin{equation}
    W_R\left( \bm{k} + \sum_{i=1}^n \bm{k}_i \right) = W_R(k) \prod_i W_R(k_i) \times \prod_{i\ne j} e^{\bm{k}_i\cdot\bm{k}_j R^2} 
\end{equation}
the contribution can be rewritten as
\begin{multline}
    \delta_g^{(n)}\delta_{g,R}^{(n + 1)} = W_R(k)\delta_L(\bm{k})\int_{\bm{k}_1\cdots\bm{k}_n} Z_n(-\bm{k}_1,\dots,-\bm{k}_n)Z_{n+1}(\bm{k}_1,\dots,\bm{k}_n,\bm{k})\prod_{i\ne j} e^{\bm{k}_i\cdot\bm{k}_j R^2}   \\ \prod_i W_R(k_i) P_L(k_i)
\end{multline}
Thus, assuming that the integral over angles yields a function that neither breaks the perturbative scaling nor the Gaussian suppression in the product of windows, we will get a controlled correction to $\delta_M^{(1)}$ proportional to the product of weighted variances. 

\begin{figure}
    \centering
    \includegraphics[width=\linewidth]{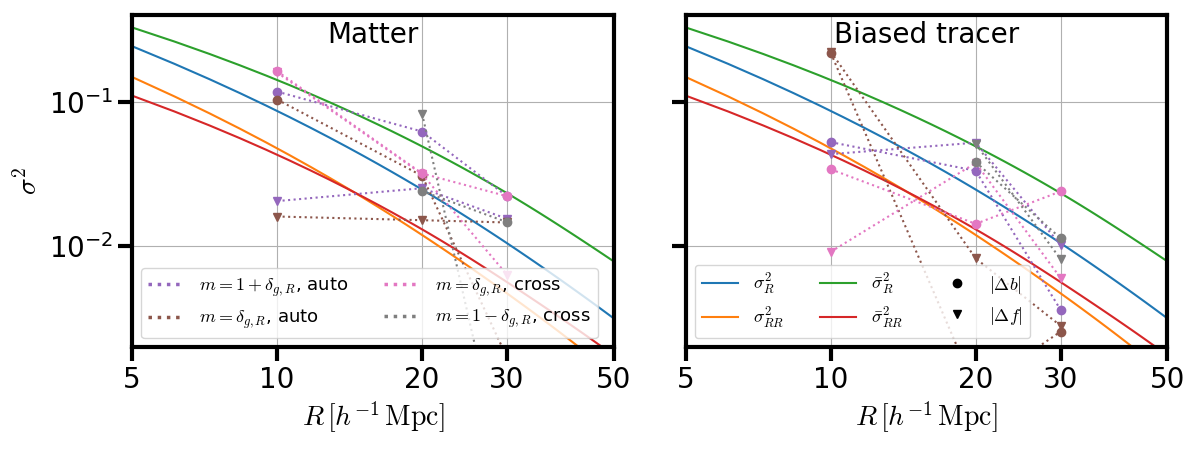}
    \caption{The smoothed variances of the density field (Eqn.~\ref{sigma_eqn}) against the low-$k$ corrections to the linear theory power-spectrum (Eqn.~\ref{lowk_correction}). As the variances are the first order contribution of the infinite series of corrections, each suppressed by further powers of these small quantities, this is a proxy for the theoretical error. Plotted against the variances are the correction parameters that we use to show the agreement between simulation and theory in \S\ref{mocks}, with the left (right) panel displaying those for the matter (biased tracer) spectrum. Note that generally the low $k$ corrections are within a factor of a few of the variances, as expected if the correction is a power series in $\sigma^2$.  Since several terms contribute, and the coefficients in the expansion are not required to be unity, we do not expect or require much better than ``order of magnitude'' agreement.}
    \label{fig:dbdf}
\end{figure}

To one-loop order the low $k$ corrections have been shown to be as described above, with perturbative corrections to the Kaiser contribution of the marked spectrum \cite{Philcox21}. This agrees with one-loop results of the LyAF correlation function as well, which encounters these corrections in UV-divergent settings \cite{Chen21}. The additional contributions to $P_L$ terms at low $k$ are proportional to four variances
\begin{align} \label{sigma_eqn}
    \sigma_R^2 = \int_{\bm{p}} W_R(p) P_L(p), &\qquad 
    \sigma_{RR}^2 = \int_{\bm{p}} W_R^2(p) P_L(p)  \\
    \bar{\sigma}_{R}^2 = \int_{\bm{p}} (pR)^2 W_R(p) P_L(p),  &\qquad
    \bar{\sigma}_{RR}^2 = \int_{\bm{p}} (pR)^2 W_R^2(p) P_L(p) 
\end{align}
where all four arise from the bottom left diagram of Figure \ref{fig:1leg1} and the top two from the bottom right diagram. As shown in Figure \ref{fig:dbdf}, these corrections are percent level for a ``reasonably sized'' smoothing radius ($R\sim 20\,h^{-1}$Mpc, so $k_R \sim 0.05\,h\,\mathrm{Mpc}^{-1}$ comparable to tree-level bispectrum analysis scales \cite{Ivanov23}).  Both from perturbative assumptions and our order $2n+1$ calculations, higher order terms are expected to contribute proportional to those variances or higher powers of them so the correction should form a converging series in $\sigma^2$.  Going forward, we will include corrections to linear theory from all orders using a function of the form
\begin{equation}
    (b+\Delta b + (f+\Delta f)\mu^2)^2 P_L
    \label{lowk_correction}
\end{equation}
to test the analytical calculations against simulations in \S\ref{mocks}. Note that inevitably we have converged to a similar correction form as with the other cosmological statistics that encounter contact terms, namely the LyAF correlation function \cite{McDonald00,McQuinn11,Chen21,Ivanov24,DESI24-IV}. However, in comparison to the UV-divergent, uncontrolled corrections encountered in the LyAF formalism, we encounter the corrections in a UV-safe, controlled environment. Placed into perspective, the corrections encountered in LyAF correlations are typically order unity \cite{deBelsunce24,Rogers18,Slosar11}, whereas the corrections to the marked formalism are percent level as shown in Figure \ref{fig:dbdf}. 

In addition, our choice of the mark and smoothing scale gives us control over this correction. 
First, as these low-$k$ corrections enter as a result of higher-loop corrections, reducing the complexity of the mark aids in its reduction. Hence, marks that expand infinitely in powers of $\delta_{g,R}$, such as that adopted in previous studies \cite{White16,Philcox21}, could be expected to increase this correction significantly.  Conversely our low-order polynomial marks should minimize this correction. 
Second, the  choice of smoothing scale is a powerful handle over the magnitude of corrections, as shown in Figure \ref{fig:dbdf}, with the correction magnitudes varying by nearly an order of magnitude between ``reasonably'' sized smoothing scales of $R=10\,h^{-1}$Mpc and $30\,h^{-1}$Mpc. 
These characteristics not only demonstrate the stability and theoretical clarity of the marked formalism, but also indicate that this is a useful theoretical `playground' for exploring the effect of these low $k$ ``biases'' common in cosmology. 

\section{Cross marked power spectra} \label{cross}

As the marked overdensity field consist of contributions linear in the mark expansion coefficients $C_n$, one can easily calculate the cross-spectrum between two marked overdensity fields by symmetrizing the coefficients as one would do for biases in a typical cross-spectrum, as long as the smoothing radius and bias parameters are shared. 
This means that Eqn.~\ref{M_A}, \ref{M_B}, and \ref{M_C} transform to the following form for the cross-spectrum of marks characterized by (\{$C_n^a$\}, \{$C_n^b$\}) (see Eqn.~\ref{Cn} for definition)
\begin{align}
\begin{split}
    M_{11} &= M_{11}^A = C_{\delta_M}^a(k) C_{\delta_M}^b(k) (k)Z_1(\bm{k})^2 P_L(k) = C_{\delta_M}^a(k) C_{\delta_M}^b(k) P_{11}(\bm{k})\\
    M_{13} &\supset M_{13}^A = 3 C_{\delta_M}^a(k) C_{\delta_M}^b(k) Z_1(\bm{k}) \int_{\bm{p}} Z_3(\bm{p},-\bm{p},\bm{k})P_L(p) = C_{\delta_M}^a(k) C_{\delta_M}^b(k) P_{13}(\bm{k}) \\
    M_{22} &\supset M_{22}^A = 2 C_{\delta_M}^a(k) C_{\delta_M}^b(k) \int_{\bm{p}} Z_2(\bm{p},\bm{k}-\bm{p})^2 P_L(p) P_L(|\bm{k}-\bm{p}|) = C_{\delta_M}^a(k) C_{\delta_M}^b(k) P_{22}(\bm{k}) 
\end{split}\\
\begin{split}
    M_{13} &\supset M_{13}^B = 2 Z_1(\bm{k})P_L(k) \int_{\bm{p}} \left[ C_{\delta_M}^a(k) C_{\delta_M^2}^b(p,|\bm{k}-\bm{p}|) + C_{\delta_M}^b(k)C_{\delta_M^2}^a(p,|\bm{k}-\bm{p}|) \right]  \\
    & \hspace{6cm} \times Z_1(\bm{p})Z_2(\bm{k},-\bm{p})P_L(p)   \\
    M_{22} &\supset M_{22}^B = 2\int_{\bm{p}} \left[ C_{\delta_M}^a(k) C_{\delta_M^2}^b(p,|\bm{k}-\bm{p}|) + C_{\delta_M}^b(k)C_{\delta_M^2}^a(p,|\bm{k}-\bm{p}|) \right] \\
    & \hspace{6cm} \times Z_2(\bm{p},\bm{k}-\bm{p})Z_1(\bm{p})Z_1(\bm{k}-\bm{p})P_L(p)P_L(|\bm{k}-\bm{p}|)    
\end{split}\\
\begin{split}
    M_{13} &\supset M_{13}^C = \frac{3}{2}  Z_1(\bm{k})^2 P_L(k) \int_{\bm{p}}\left[C_{\delta_M}^a(k) C_{\delta_M^3}^b(p,p,k) C_{\delta_M}^b(k) C_{\delta_M^3}^a(p,p,k)\right]Z_1(\bm{p})Z_1(-\bm{p}) P_L(p)  \\
    M_{22} &\supset M_{22}^C = \int_{\bm{p}} C_{\delta_M^2}^a(p,|\bm{k}-\bm{p}|)C_{\delta_M^2}^b(p,|\bm{k}-\bm{p}|) \left[Z_1(\bm{p})Z_1(\bm{k}-\bm{p})\right]^2 P_L(p)P_L(|\bm{k}-\bm{p}|) 
\end{split}
\end{align}
This indicates that by choosing specific pairs of marks, one can cancel some contributions, which may be helpful when looking for specific parameter dependencies. 
A straight-forward example of this is when the two marks share a smoothing scale and the contributions are proportional to $C_n^a C_n^b$, such that for $n\ne0$, one can eliminate these terms by crossing with the unmarked density field; e.g.~
\begin{align}
\begin{split}
    \frac{M_{13}^B}{ 2 Z_1(\bm{k})P_L(k)} &= \int_{\bm{p}} \left[ C_{\delta_M}^a(k) C_{\delta_M^2}^b(p,|\bm{k}-\bm{p}|) + C_{\delta_M}^b(k)C_{\delta_M^2}^a(p,|\bm{k}-\bm{p}|) \right]   Z_1(\bm{p})Z_2(\bm{k},-\bm{p})P_L(p) \\ 
    &\supset  C_1^a C_1^b W_R(k)\int_{\bm{p}} [W_R(p)+W_R(|\bm{k}-\bm{p}|) ]Z_1(\bm{p})Z_2(\bm{k},-\bm{p})P_L(p) \\
    &+  \frac{1}{2}[C_0^a C_1^b + C_1^a C_0^b] \int_{\bm{p}} [W_R(p)+W_R(|\bm{k}-\bm{p}|) ]Z_1(\bm{p})Z_2(\bm{k},-\bm{p})P_L(p)
\end{split}
\end{align}
where the first contribution is $\propto C_1^a C_1^b$. 
The second term is another example of a contribution that can be eliminated by tuning the mark, such that $C_0^a C_1^b + C_0^a C_1^b = 0$.
When two different tracers are used for the cross-spectrum, the situation is more involved, as contributions to $\delta_M$ consist of products of $C_n$ and biases. While it is straight-forward to calculate, we will omit the expression from this work due to its length.  

Similar to the cross-spectrum between distinct galaxy tracers in multi-tracer cosmology \cite{Mergulhao23, Ebina24}, or density-split statistics \cite{Gruen16,Friedrich18}, the cross marked power spectrum, between marks of different expansion coefficients and smoothing scales, may also improve constraints through further degeneracy breaking. Practically, we may hope that the cross-spectrum between the unmarked field and marked field yields better agreement between simulations and theory, at the possible cost of less genuinely new information.

\section{Validation against mocks} \label{mocks}

We will validate our calculations and set the range of validity using a set of mock catalogs.  These mock catalogs are built upon halo catalogs from the {\sc AbacusSummit} cosmological N-body simulation suite \cite{Maksimova21}, produced with the {\sc Abacus} N-body code \cite{Garrison18,Garrison21}.  We use 25 simulations of the $\Lambda$CDM family, each employing $6912^3$ dark matter particles in a periodic box of side length $2\,h^{-1}$Gpc \cite{Maksimova21}.  The simulations all assume the same $\Lambda$CDM cosmological parameters as Planck \cite{PCP18} ($\Omega_M=0.3153$, $h=0.6736$, $\sigma_8=0.808$), differing only in the realization of the initial conditions.  In each case we look at the dark matter field and a set of mock galaxies that are placed in halos using the AbacusUtils\footnote{https://abacusutils.readthedocs.io/en/latest/} software.

We use the outputs at $z\simeq 1.1$ and move the objects into redshift space using the velocity field of the simulation assuming the distant observer approximation, so that the line of sight is parallel to one of the simulation axes.  Figure \ref{fig:deltaR} shows the (normalized) distribution of the smoothed overdensity field in the simulations, for both matter and the biased tracer, for three different smoothing scales: $R=10$, 20 and $30\,h^{-1}$Mpc.
With these smoothing radii, the 95\% ranges in the mark $1 + \delta_{g,R}$ for matter are $[0.57,1.65]$, $[0.75,1.31]$, and $[0.84,1.19]$, respectively. Equivalently for biased tracers they are $[0.47,1.74]$, $[0.70,1.35]$, and $[0.80,1.22]$. 

If we were fitting to data we would simultaneously fit to the marked and unmarked spectra varying all of the parameters.  For the purposes of showing that the model provides good fits to the simulation data we take a different approach.  As the power spectrum is better constrained and parameters (such as the biases and counterterms) are shared between the unmarked and marked power spectrum, these nuisance parameters are first fit for using the power spectrum. Then, the additional stochastic degrees of freedom (see \S\ref{small-scale} and \S\ref{app:stochastic}) and low-$k$ corrections (Eqn.~\ref{lowk_correction}) are introduced and fit to the marked power spectrum. 

To explore what marks can be modelled using the current framework, we consider several functional forms. In the limit that the mark is unity, we will recover the power spectrum, which can be modelled extremely well \cite{Maus24}. 
On the contrary, when the mark is $m=\delta_{g,R}$, the marked spectrum at small scales ($k \gtrsim R^{-1}$) will only include terms that are integrations over three or four distinct overdensity fields, which is expected to be more difficult to model. This mark is also numerically challenging to measure in simulations, as for $R$ too large, the signal tends to 0.  The situation worsens when $m=\delta_{g,R}^2$.  Thus, we will expect the limitations of modelling to lie somewhere between these marks.
Specifically, we test marks of the form $m=\{0, 1\} \pm \delta_{g,R}$. While these marks are in-principle degenerate \cite{Cowell24}, as they only vary by constant additions to the mark, limiting the dynamic range of mark can be beneficial for both better numerical behavior and a more Gaussian covariance.  It should also help when applied to real data which can additionally contain systematic and observational artifacts -- reducing the dynamic range of the mark avoids overweighting any particular region of the survey. Having a larger constant contribution of the mark also protects the covariance from being too non-Gaussian \cite{Cowell24} and being overly sensitive to finite survey volume.  The downside of introducing the constant is that it makes $M$ more covariant with $P$.

A recent study investigating the possibility of an optimal mark using N-body simulations \cite{Cowell24} has reported that the optimal mark has a shape peaking at highly under- or over-dense regions. While we have explored similar regimes through modelling of quadratic marks, we have found the agreement between theory and simulations to be poor. 

\subsection{Matter} \label{matter}

\begin{figure}
    \centering
    \includegraphics[width=\linewidth]{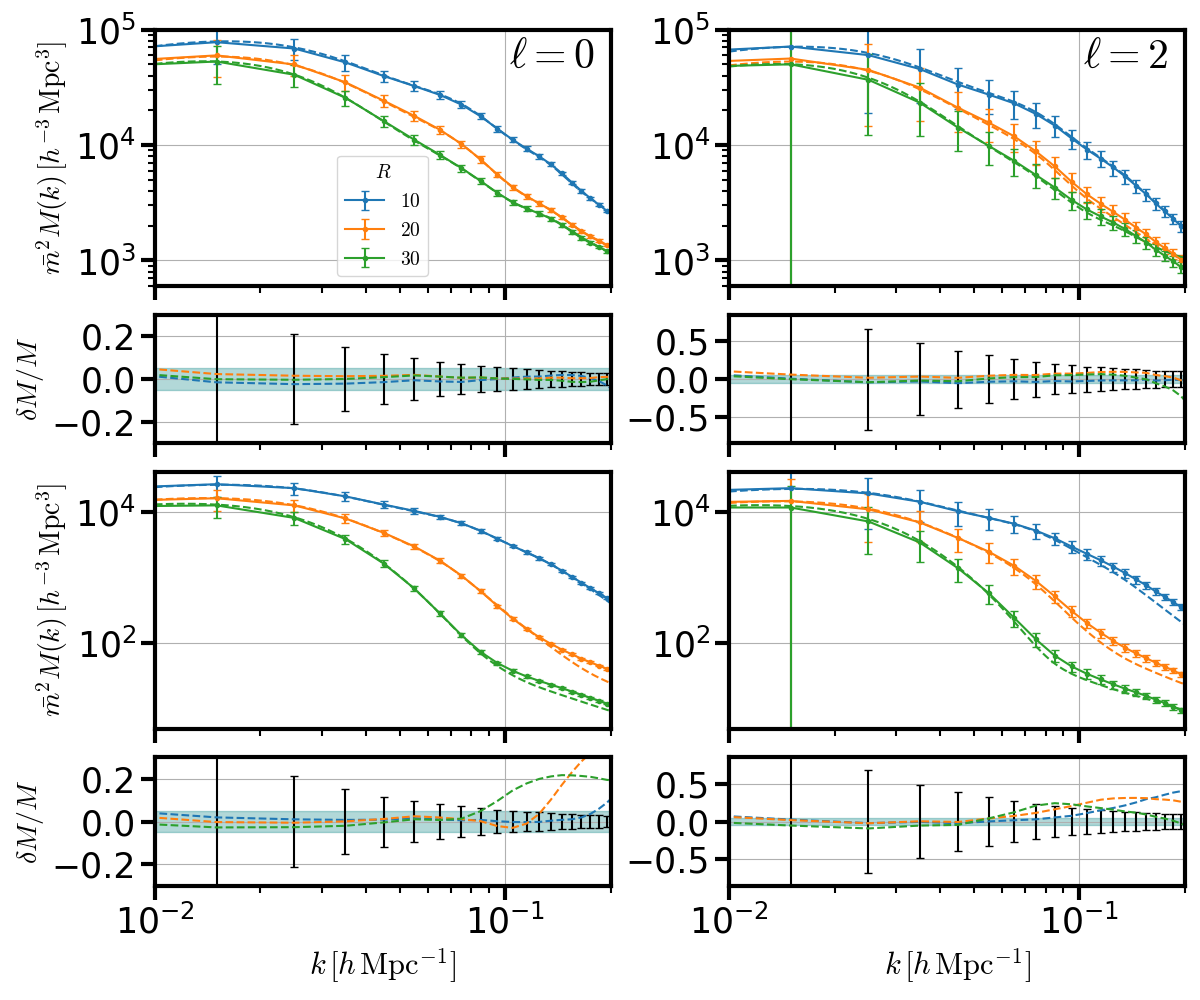}
    \caption{Comparison between the analytical calculations and Abacus N-body simulations for the auto-spectrum of mark $m=1+\delta_{g,R}$ (top) and $m=\delta_{g,R}$ (bottom) for smoothing radii $R=\{10, 20, 30\}\,h^{-1}$Mpc for the matter. The presence of the constant term in the mark $m=1+\delta_{g,R}$ limits the dynamic range, which has observational benefits, but also makes $M(k)$ less independent of $P(k)$. The shaded bands in each residual plot is showing 5\% range to aid visualization.}
    \label{fig:matter_auto}
\end{figure}

First we will test the theory against the matter field.  We will explore Gaussian smoothing radii of $R=10$, $20$ and $30\,h^{-1}\mathrm{Mpc}$ for $\delta_{g,R}$ and several functional forms for the mark.  The comparison for the auto-spectra of marks $m=1+\delta_{g,R}$ and $m=\delta_{g,R}$ between simulations and theory are shown in Figure \ref{fig:matter_auto}. 
In the case of $m=1+\delta_{g,R}$ we see a good agreement between simulations and theory (within $1\,\sigma$) up to $k\simeq 0.2\,h\,\mathrm{Mpc}^{-1}$, which is the max extent of our fitting range.  We have not binned the theory into the same $k$ bins as the numerical data, which results in a small correction at low $k$ especially when $M$ is changing rapidly with $k$.  Since this region is not our main focus the small error this introduces will not be of concern.

The case of $m=\delta_{g,R}$ is more challenging, as expected. We observe that there is agreement (within $1\,\sigma$) until $k\simeq 0.08\,h\,\mathrm{Mpc}^{-1}$ for all smoothing radii, which is similar to the $k_{\rm max}$ adopted for the tree-level bispectrum in ref.~\cite{Ivanov19}. For $R=10\,h^{-1}$Mpc, we see that the agreement continues to higher $k$. This may seem surprising as the intuition is that we will be able to model a strongly smoothed (and as a result more Gaussian) field better. However, recall that the power-spectrum contribution to the marked power spectrum scales as $C_{\delta_M}^2(k)$ (Eqn. \ref{M_A}). For this mark, $m=\delta_{g,R}$, this is $W_R^2(k)$, which is significantly larger at $k>0.1\,h\,\mathrm{Mpc}^{-1}$ for $R=10\,h^{-1}$Mpc than the other smoothing radii we consider. In particular, at $k=0.1\,h\,\mathrm{Mpc}^{-1}$, the power-spectrum contribution to the marked spectrum is a factor of $\sim20$ ($\sim3000$) greater with a smoothing radii of $R=10\,h^{-1}$Mpc than with $R=20$(30)$h^{-1}$Mpc, highlighting the ease of modeling the $R=10\,h^{-1}$Mpc mark at smaller scales. 

\begin{figure}
    \centering
    \includegraphics[width=\linewidth]{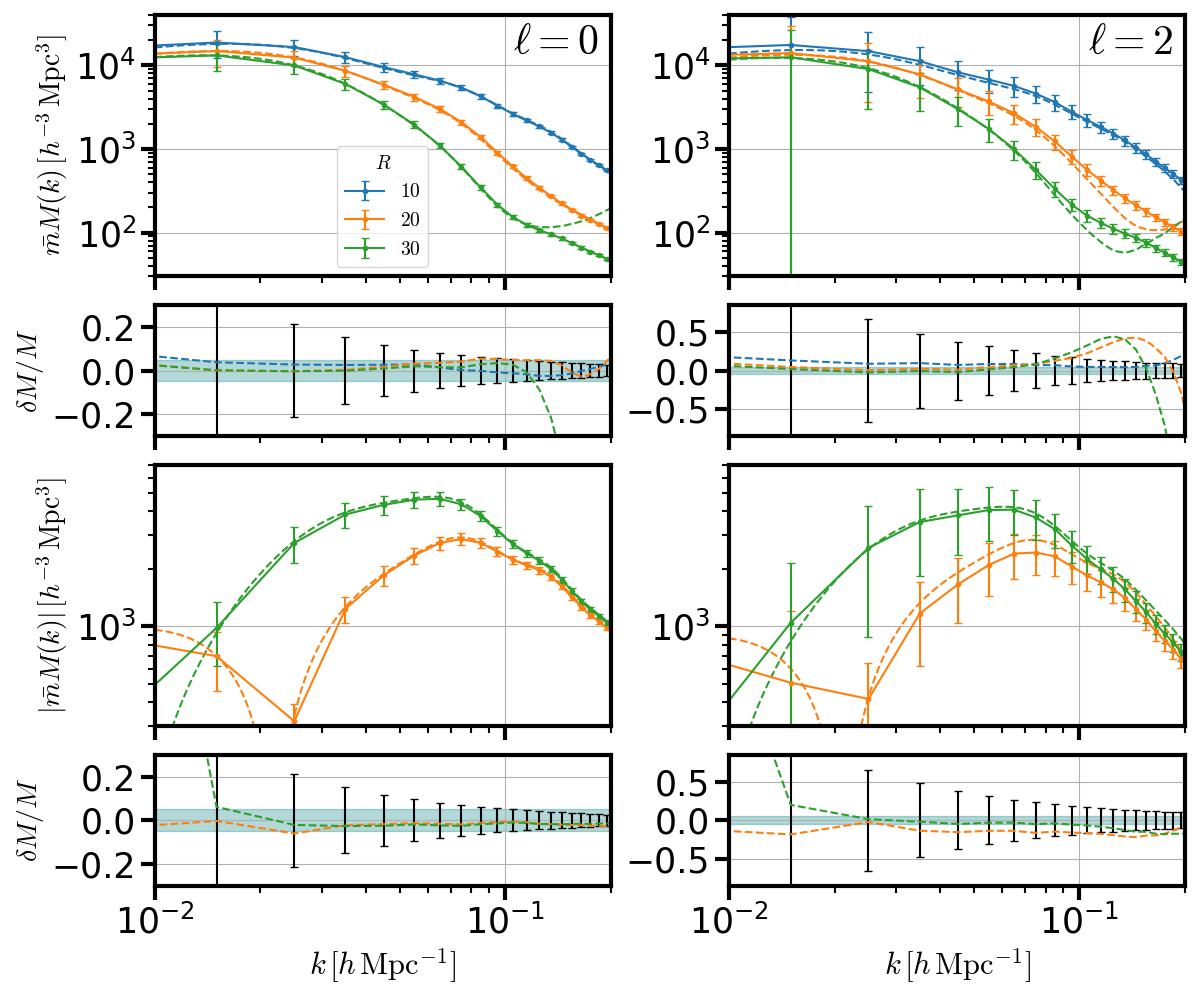}
    \caption{Comparison between the analytical calculations and Abacus N-body simulations for the cross-spectrum between the unmarked field $m=1$ and marked fields for the matter. The marked fields used are $m=\delta_{g,R}$ (top) and $m=1-\delta_{g,R}$ (bottom). While for $m=\delta_{g,R}$ results are shown for smoothing radii $R=\{10, 20, 30\}\,h^{-1}$Mpc, for the $m=1-\delta_{g,R}$ the $R=10\,h^{-1}$Mpc results are suppressed as $\delta_{g,R}>1$ for a non-negligle portion of the population for a $R=10\,h^{-1}$Mpc, as shown in Figure \ref{fig:deltaR}, making the evaluations of the marked spectrum numerically difficult.  As discussed in the text, the theory is not binned in $k$ bins, and so can deviate from the (binned) N-body points in regions where $M$ is changing rapidly. }
    \label{fig:matter_cross}
\end{figure}

As described in \S\ref{cross}, the cross-spectrum between marked fields is an interesting direction to explore. In particular, the cross-spectrum between the unmarked field and marked field $\delta_{g,R}$ may be easier to model than the auto-spectrum due to good behavior of perturbative models of the unmarked field. 
In Figure \ref{fig:matter_cross} we show the comparison for this cross-spectrum.  We remind the reader that the theory curves are not binned to match the N-body, which explains the small deviations at low $k$ where $M$ is changing rapidly.  Since we have theoretical reasons to believe our model for very low $k$ we have not attempted to correct for this.  Bearing that in mind, we observe that there is agreement up to $k\sim0.08\,h\,\mathrm{Mpc}^{-1}$ for all smoothing radii, as expected from the $k_{\rm max}$ adopted for the tree-level bispectrum \cite{Ivanov19}. Although the $k$-range of 1$\sigma$ agreement is similar to the auto-spectrum, the residuals in the cross-spectrum have less structure, suggesting that it may be easier to model. 

Finally we check our modelling consistency for a specific mark that we find particularly useful in terms of degeneracy breaking, as discussed later in \S\ref{degeneracy}. The mark is $m=1-\delta_{g,R}$, and as we find that its cross-spectrum with the unmarked field is sufficient for degeneracy breaking, we will present the comparison in Figure \ref{fig:matter_cross}. We omit the $R=10\,h^{-1}$Mpc curve from the figure, due to the large dynamic range of the mark with the mark crossing zero in denser regions (see the positive tail of $\delta_{g,R}$ in Figure \ref{fig:deltaR}).  For larger smoothing radii, the dynamical range of the mark is considerably restricted and we find $1\,\sigma$ consistency beyond $k\simeq 0.1\,h\,\mathrm{Mpc}^{-1}$, where the model begins to deviate slightly. 

The left panel of Figure \ref{fig:dbdf} shows low-$k$ corrections from higher loops that we have fitted for in order to obtain the curves above. We see that the corrections ($\Delta b$ and $\Delta f$ described in Eqn.~\ref{lowk_correction}) match the level expected from a power series in the smoothed variances of the density field. 

The errorbars quoted in the Figures above are calculated for the volume of one simulation box ($8\,h^{-3}\mathrm{Gpc}^3$) assuming the disconnected piece of the covariance dominates. For large scales this should be a reasonable approximation, and the volume is roughly comparable to that probed by the current generation of large redshift surveys.  While this is clearly not a complete treatment, this will serve our needs here which are primarily for visualization purposes. We have checked the variance against the scatter of 25 independent simulation boxes. While the scatter reveals a non-Gaussian distribution of $M(k)$, indicating a need for an improved covariance calculation in the future, we find acceptable agreement for our purposes. Similarly for visualization purposes, in the lower panels we have shown the error bars appropriate to one smoothing scale, taken to be $R=30\,h^{-1}$Mpc, even though we show curves for multiple $R$.  In the sample variance limit the fractional error is independent of $R$, and we find our error estimates are only weakly dependent on $R$ for the relevant scales.  Ref.~\cite{Cowell24} have looked at other contributions to the covariance using analytic and numerical approaches. They show that the first correction is of low-rank ($\propto P(\bm{k})P(\bm{k'})$) and thus unlikely to significantly impact constraints on parameters -- we believe that excluding this will not change our conclusions dramatically.

\subsection{Biased Tracers}

Application of the marked power spectra to data would most likely occur for biased tracers such as galaxies or QSOs rather than for the matter field.  The presence of bias terms leads to a different weighting of the contributions to our theoretical model, and could change the level of agreement between theory and simulation.  For this reason we also compare our theoretical model to catalogs of mock galaxies in redshift space.  Our mock galaxies are generated at redshift $z=1.1$ from an HOD model using the AbacusHOD software \cite{abacusHOD}.  The HOD is of a standard form \cite{Zheng11} with parameters $\log_{10} M_{\rm cut}=11$ and $\log_{10} M_1=12.3$.  These were adjusted\footnote{We also investigated samples with a higher bias and lower number density, obtaining qualitatively similar results.} to give a linear bias $b= 1.2$ and also happen to lead to quadratic bias $b_2=-1.8$ and shear bias $b_s=0.5$.  For completeness the best-fit values of the stochastic terms are $N = 35\,h^{-3}\mathrm{Mpc}^3$ and $N_2 = -4.7\times10^3\,h^{-5}\mathrm{Mpc}^5$, and the counterterms $\alpha_0 = -0.4\,h^{-2}\mathrm{Mpc}^2$ and $\alpha_2 = 6.6h^{-2}\mathrm{Mpc}^2$. 

\begin{figure}
    \centering
    \includegraphics[width=\linewidth]{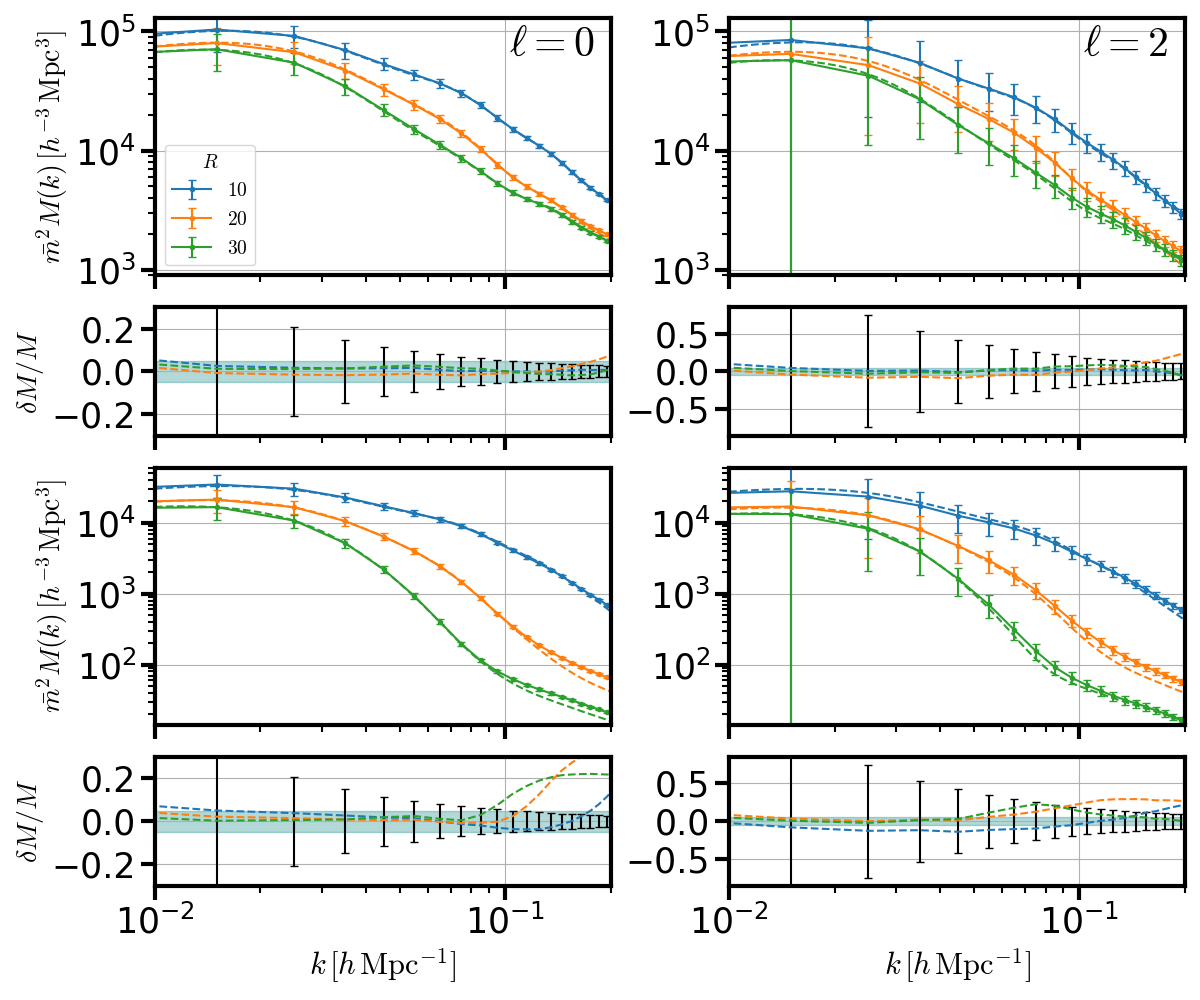}
    \caption{Comparison between the analytical calculations and AbacusHOD biased tracers for the auto-spectrum of mark $m=1+\delta_{g,R}$ and smoothing radii $R=\{10, 20, 30\}\,h^{-1}\mathrm{Mpc}$. The presence of the constant term in the mark limits the dynamic range, which has observational benefits, but also makes $M(k)$ less independent of $P(k)$. Refer to Figure \ref{fig:matter_auto} for the matter field agreement using the same mark.
    }
    \label{fig:LRG_auto}
\end{figure}

\begin{figure}
    \centering
    \includegraphics[width=\linewidth]{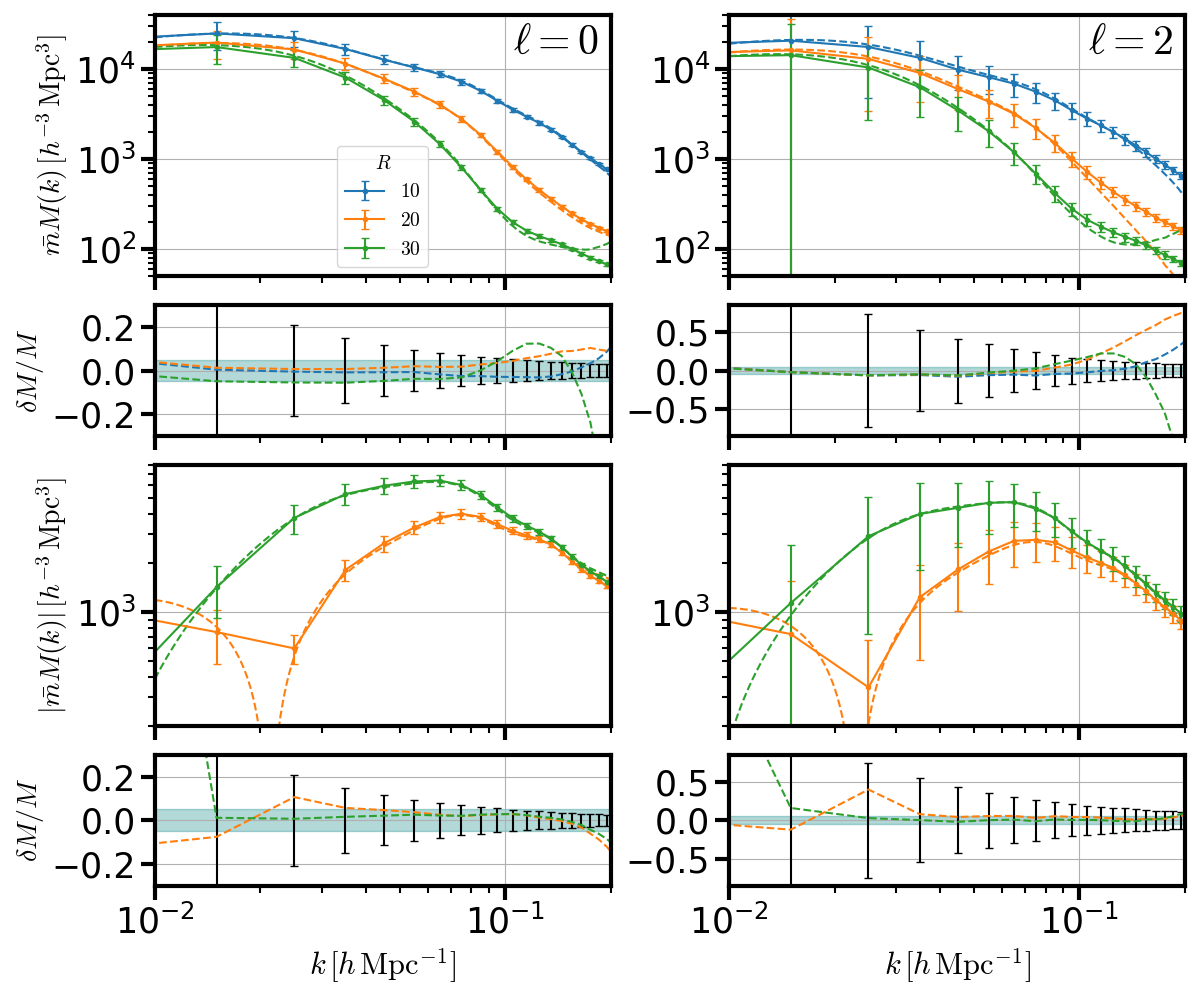}
    \caption{Comparison between the analytical calculations and AbacusHOD biased tracers for the cross-spectrum between the unmarked field $m=1$ and marked fields. The marked fields used are $m=\delta_{g,R}$ (top) and $m=1-\delta_{g,R}$ (bottom). While for $m=\delta_{g,R}$ results are shown for smoothing radii $R=\{10, 20, 30\}\,h^{-1}$Mpc, for the $m=1-\delta_{g,R}$ the $R=10\,h^{-1}$Mpc results are suppressed as $\delta_{g,R}>1$ for a non-negligle portion of the population for a $R=10\,h^{-1}$Mpc, as shown in Figure \ref{fig:deltaR}, making the evaluations of the marked spectrum numerically difficult. Refer to Figure \ref{fig:matter_cross} for the matter field agreement using the same marks.
    }
    \label{fig:LRG_cross}
\end{figure}

In Figures \ref{fig:LRG_auto} and \ref{fig:LRG_cross} we show results paralleling those for the unbiased, matter field in \S\ref{matter} above. Once again, the combination of marks probed are the auto-spectrum of $m=1+\delta_{g,R}$, auto-spectrum of $m=\delta_{g,R}$, cross-spectrum between the unmarked field and $m=\delta_{g,R}$, and cross-spectrum between the unmarked field and $m=1- \delta_{g,R}$. 

In general, we find similar results with the matter case. 
For the auto-spectra shown in Figure \ref{fig:LRG_auto}, there is agreement to near $k=0.2\,h\,{\rm Mpc}^{-1}$. 
The auto-spectrum of $m=\delta_{g,R}$ again find agreement up to $k=0.09\,h\,{\rm Mpc}^{-1}$, as expected from tree-level bispectrum analysis limits \cite{Ivanov19}.

The cross-spectrum between the unmarked field and $m=\delta_{g,R}$ shown in Figure \ref{fig:LRG_cross}, we observe that there is better agreement than the auto-spectrum in the theoretically expected regime $k<0.08\,h\,{\rm Mpc}^{-1}$, with agreement within 5\% for all smoothing radii, and a slightly higher 1$\sigma$ agreement until $k=0.1\,h\,{\rm Mpc}^{-1}$. 
Finally, the situation is similar with the matter case for the cross-spectrum between the unmarked field and $m=1-\delta_{g,R}$, shown in the bottom panels of the same figure. There is 1$\sigma$ agreement for almost the entire $k$-range for both smoothing radii. Similar to the matter case the $R=10\,h^{-1}\,{\rm Mpc}$ is suppressed due to the dynamic range of the mark (see Figure \ref{fig:deltaR}). 
As we will see soon, this agreement will be important to potentially exploit the degeneracy breaking of the marked power spectrum.

\section{Degeneracy breaking} \label{degeneracy}

The major reason to consider additional statistics beyond $P(k)$ is to see whether we can break degeneracies between the many parameters in our models that otherwise limit our constraints on cosmological parameters of interest.  For example, it is known that the bispectrum aids in breaking degeneracies involving $b_2$ and thus improves cosmological constraints \cite{Gil-Marin16,Ivanov23}. As the marked power spectrum includes integrals of the bispectrum, it is natural to investigate its potential to break the same degeneracy. 

\begin{figure}
    \centering
    \includegraphics[width=\linewidth]{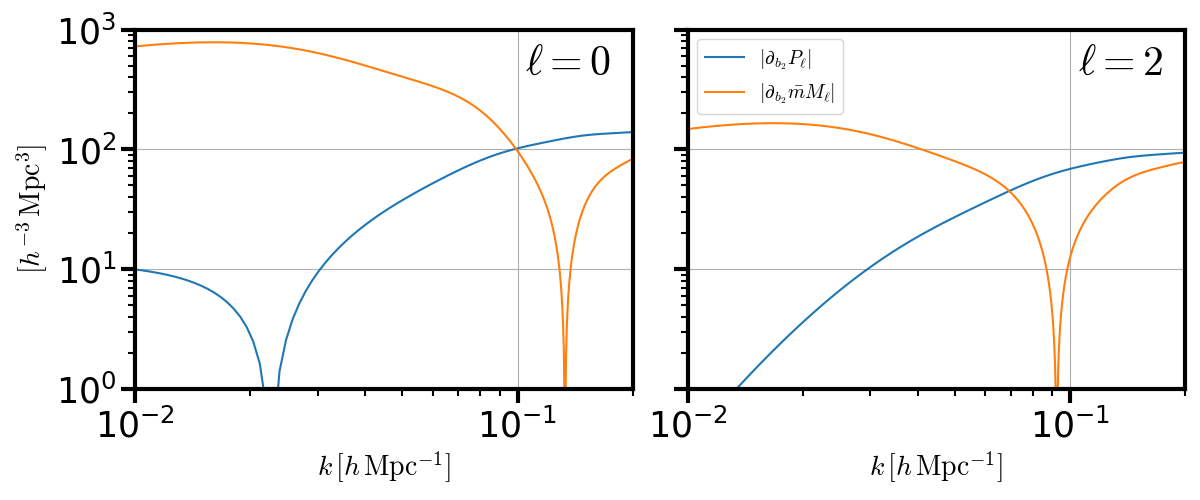}
    \caption{A comparison of the $b_2$ dependence of the power spectrum (blue lines), and the cross-marked power spectrum between the unmarked field $m=1$ and marked field $m=1-\delta_{g,R}$ (orange lines). The absolute value of each curve is shown due to their crossing of 0. As expected from the analytical calculation there is a different scale dependence, raising the possibility of using $M(k,\mu)$ for degeneracy breaking.
    }
    \label{fig:db2}
\end{figure}

At the equation level, it is easy to isolate contributions to the marked power spectrum including $b_2$, as this only arises from terms that involve the perturbative redshift-space kernels $Z_2$ and $Z_3$, which means that we focus on terms involving $\delta_g^{(2)}$'s and $\delta_g^{(3)}$'s in Figure \ref{fig:diagrams}. In particular, they are both linear in $b_2$
\begin{align}
    Z_2(\bm{k_1}, \bm{k_2}) &\supset \frac{b_2}{2} \\ 
    Z_3(\bm{k_1}, \bm{k_2}, \bm{k_3}) &\supset [b_2 F_2(\bm{k_1},\bm{k_2}) + {\rm cycl.}]
\end{align}
where $F_2$ is the second-order density kernel in real-space \cite{Bernardeau02}.
One set of contributions are those included in the ``two-point'' contribution $M_{13}^A$ and $M_{22}^A$ (Eqn.~\ref{M_A}). These are simply the power-spectrum dependence of $b_2$ times an overall factor $C_{\delta_M}^2(k)$. 
The others are those coming from ``three-point'' terms $M_{13}^B$ and $M_{22}^B$ (Eqn.~\ref{M_B}).
The $b_2$ dependence can then be isolated as
\begin{align}
    \frac{1}{2 C_{\delta_M}(k)} \frac{\partial}{\partial b_2} M_{13}^A &= 3C_{\delta_M}(k) P_L(k) \int_{\bm{p}} F_2(\bm{k},\bm{p}) P_L(p) \\
    \frac{1}{2 C_{\delta_M}(k)} \frac{\partial}{\partial b_2} M_{22}^A &= C_{\delta_M}(k) \int_{\bm{p}} Z_2(\bm{p},\bm{k}-\bm{p})  P_L(p) P_L(|\bm{k}-\bm{p}|)\\
    \frac{1}{2 C_{\delta_M}(k)} \frac{\partial}{\partial b_2} M_{13}^B &= Z_1(\bm{k}) P_L(k) \int_{\bm{p}} C_{\delta_M^2}(p,|\bm{k}-\bm{p}|) Z_1({\bm{p}}) P_L(p) \\
    \frac{1}{2 C_{\delta_M}(k)} \frac{\partial}{\partial b_2} M_{22}^B &= \int_{\bm{p}} C_{\delta_M^2}(p,|\bm{k}-\bm{p}|) Z_1({\bm{p}}) P_L(p) Z_1(\bm{k}-\bm{p}) P_L(|\bm{k}-\bm{p}|) 
\end{align}
suggesting that the $b_2$ dependence of the marked and unmarked power spectrum are indeed different. This is confirmed in Figure \ref{fig:db2}.

\begin{figure}
    \centering
    \includegraphics[width=\linewidth]{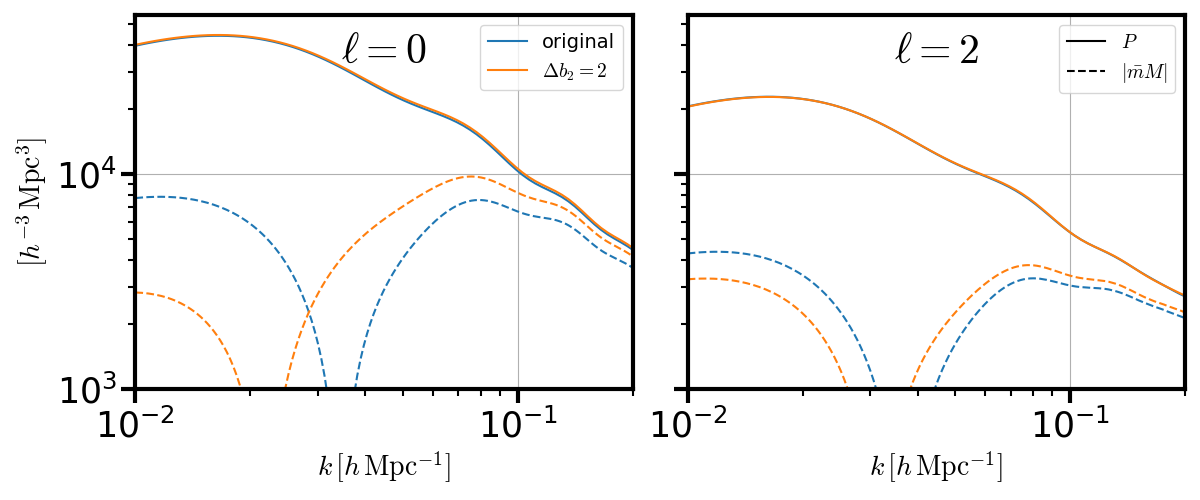}
    \caption{Comparison of spectra between two bias parameter sets that differ in $b_2$ by 2 (with other nuisance parameters adjusted to keep the power spectrum almost unchanged). It is visually clear that the power spectra (solid lines) for the two parameter sets are almost indistinguishable while the marked spectra (dashed lines) differ significantly. In this particular example $M$ is the cross-spectrum between the unmarked field ($m=1$) and $m=1-\delta_{g,R}$, with a smoothing radius $R=20\,h^{-1}$Mpc.}
    \label{fig:degen_comparison}
\end{figure}

Given that there is a visually apparent difference in parameter dependence, whether this difference is practically large enough for us to exploit in the near-future is an important question. While we will leave a complete investigation for a future study, we will test its potential by using two parameter sets that are degenerate to the power spectrum and showing that they are distinguishable using the marked power spectrum. 
We find and select the parameter sets such that their $b_2$ differ by 2 in order to exploit the difference in $b_2$ dependence found above. $\Delta b_2\sim2$ is not unreasonable for one redshift bin for current generation surveys \cite{Maus24}. For $m=1-\delta_{g,R}$ and $R=20\,h^{-1}$Mpc, Figure \ref{fig:degen_comparison} shows that $M$ is significantly different for the two sets of parameters, despite them being nearly degenerate in $P$. 
With respect to errors computed through Gaussian covariance approximations, at $k\sim0.05\,h\,{\rm Mpc}^{-1}$ (near the regime where we expect the most theoretical constraining power) this corresponds to a SNR of 0.23 and 6.0 for $P$ and $M$, respectively\footnote{As predicted earlier in the text, we have confirmed that a simulation-based covariance computed through 25 ``base'' (2$h^{-1}$Gpc) Abacus simulations returns results consistent with the Gaussian approximation up to a factor of $\sim 2$}. Even without exploiting the difference in scale dependence of the signals, there is a gain in sensitivity by a factor of $\sim 25$, showing the effectiveness using the marked spectrum. 

By choosing $R=20\,h^{-1}$Mpc when defining $m(\delta_{g,R})$ we have ensured that the low $k$ corrections from higher order terms remain relatively small.  This is important for a mark such as $1-\delta_{g,R}$ as the linear theory contributions at low $k$, that ``stabilize'' our theory, are suppressed.  Decreasing $R$, and thus increasing the low $k$ corrections, can worsen agreement between theory and simulation, as we have noted in Figures \ref{fig:matter_cross} and \ref{fig:LRG_cross} for $R=10\,h^{-1}$Mpc.

\section{Conclusion} \label{conclusion}

An advantage of perturbative models of large-scale structure is that they rely on only a minimal set of theoretical assumptions to model a wide range of clustering data, and they produce robust and reliable inferences.  A drawback is that they tend to have a large number of parameters and, unfortunately, some of these can be (partially) degenerate with cosmological parameters.  In order to break these degeneracies additional data or additional statistics need to be introduced.  A popular set of statistics are those beyond the 2-point function, including for example the bispectrum.  Inclusion of the bispectrum, using the same basic perturbative formalism as used for the power spectrum, is known to help break degeneracies.  However the bispectrum has technical difficulties associated with it, including the measurement and sensitivity to survey systematics, handling the survey geometry (window functions) and covariance, and the size of the data vector.  Alternatives to the bispectrum include statistics involving products of fields, such as marked power spectra.  For marked spectra in particular, the connection with higher-order functions is relatively straightforward to see \footnote{After our work was finished ref.~\cite{Marinucci24} appeared which also notes the connection between marked spectra and higher-order functions and the role they can play in degeneracy breaking.} and much of the survey infrastructure developed for the power spectrum can be reused with minimal modifications. While this does not alleviate all of the technical difficulties associated with higher-point functions, it renders the marked power spectra an attractive target for further theoretical attention.

In this work we have investigated the applicability of perturbative methods for modeling the marked power spectrum for marks that are low-order polynomials in the smoothed overdensity field. This differs from previous works that focused on an inverse-density-marked power spectrum \cite{White16,Philcox21}. While this older mark has an infinite Taylor expansion in terms of the smoothed overdensity field, we consider a simpler, low-order mark.  Our intuition for this is that $\delta_{g,R}$ probes $k$ modes up to $\sim R^{-1}$ while $\delta_{g,R}^n$ probes modes up to $\sim n\,R^{-1}$.  Any products of $\delta_{g,R}$ with itself or with $\delta$ thus increase the high-$k$ sensitivity more for large $n$ than for small $n$.  Alternatively multiplication by a high power of $\delta_{g,R}$ upweights the tails of the distribution, where we expect perturbative methods to fare less well.  Setting the mark to a low-order polynomial thus aids in theoretical control and we have shown improves consistency with simulations.

We build upon the work of \cite{Philcox20,Philcox21} and present a full 1-loop Eulerian perturbation theory model (including terms to account for small-scale physics that we do not explicitly model) for the marked power spectra of biased tracers in redshift space.  This involves the characterization of the marked field using the Taylor expansion coefficients (which for our polynomial mark is exact), the inclusion of all dynamical contributions to one-loop order in the power spectrum, as well as higher-loop contributions that influence the linear-order power spectrum non-negligibly. 

One feature of the model that does not appear in the standard power spectrum is higher loop-order corrections that persist even at low $k$.  These terms have been identified in previous work \cite{Philcox21}.  We have shown, both analytically and through simulations, that these corrections will enter in a theoretically well-controlled way. Such large-scale corrections are known to the community and have been seen in other settings that deal with fields that are non-linear in the density field, e.g.\ the Lyman-$\alpha$ flux two-point function \cite{McDonald00,McQuinn11,Chen21,Ivanov24,DESI24-IV}. However, the marked power spectrum is interesting in that the corrections are UV-safe and can be controlled perturbatively.  This is because the contributing terms involve only one unsmoothed field, so the general phenomenon can be studied in a theoretically well-understood manner. 

We have investigated the potential of a cross-spectrum between two marked density fields as well. Because the marked field is a low-order polynomial in the Taylor expansion coefficients of the mark, this is a simple extension from the auto-spectrum as long as the biases are shared.  We pay particular attention to the interesting case of the cross-correlation of a marked and unmarked field.

Testing against simulations, we have found good consistency between simulation and theory for our preferred marks, though we struggle more with marks that have larger dynamic range and probe further into the tails of the density distribution. Importantly, we have demonstrated in the simulations that the low-$k$ corrections to linear theory are indeed small and agree with our theoretical expectations. 

We have also demonstrated that the marked power spectrum is capable of breaking degeneracies of the power spectrum, e.g.\ providing another route to constraining the scale-dependent bias, $b_2$.  We show explicitly how degeneracy breaking applies to the $b_2$ term, both through calculations and through a realistic example. This result is not unexpected, as it is known that the bispectrum can break degeneracies involving $b_2$ and the marked power-spectrum includes contributions from the bispectrum. Nonetheless this validates our expectation that marked power spectra could be useful for cosmological inference. 

While this work encourages the further study of the marked power spectrum, there are necessary future steps before this formalism can be applied to data. 
The treatment of Alcock-Pacynski (A-P) effects is an important one, as this enters at the very beginning of this framework when defining the smoothed overdensity fields that the mark is a function of.  We would need to generalize our treatment to anisotropic smoothing filters or handle this at the level of the data.  We have also neglected the impact of survey boundaries or holes on the computation of the smoothed density field, or the practicalities of implementing the smoothing, and the impact of shot noise \cite{Karcher24}.  The covariance matrix of the marked power spectrum is another open question. For this work, for visualization purposes, we have adopted an approximate covariance that assumes $\delta_M$ is Gaussian. While this is a reasonable approximation for large scales and large $R$, it is clearly an idealization.  Recent work has argued that the first correction to this approximation involves a low-rank matrix, driven by the stochasticity of the mark \cite{Cowell24}.  We believe such corrections are unlikely to change our main conclusions regarding degeneracy breaking (a point also emphasized in \cite{Cowell24}).

At the theoretical level, we have also identified the potential for new stochastic terms in the marked power spectrum. While we leave detailed study for the future, we have shown that a symmetry-limited ansatz suffices for satisfactory agreement between theory and simulations. 
Likewise, the correct treatment of IR-resummation beyond the first order approximation of its effect on the terms proportional to the power spectrum requires future work. Similar to the case of stochastic terms, we have identified that a correction to this approximation is not immediately required to match simulation results.
We hope to return to these questions in future work. 

\section{Data Availability}

Software used for the analysis and making of figures in this work are publicly available at \href{https://github.com/HarukiEbina/markedPS}{\texttt{https://github.com/HarukiEbina/markedPS}}. 

\section*{Acknowledgements}
We thank Stephen Chen and Zvonimir Vlah for helpful discussions pertaining to this work.  The authors are supported by the DOE.
This research has made use of NASA's Astrophysics Data System and the arXiv preprint server.
This research used resources of the National Energy Research Scientific Computing Center (NERSC), a Department of Energy Office of Science User Facility.

\appendix
\section{Stochastic corrections}
\label{app:stochastic}

This appendix presents more details on the stochastic contributions that can contribute to the marked power spectrum, and the subset of terms that we keep in our model when comparing to N-body simulations.  As discussed in the main text there are stochastic terms from the power spectrum
\begin{equation}
    \bar{m}^2M(k,\mu) \supset C_{\delta_M}^2(k) P_{st.} = C_{\delta_M}^2(k) (N + N_2 k^2 \mu^2)
    \label{M_A_correction_appendix}
\end{equation}
including the shot-noise contribution and first-order Finger of God correction. 

For the three-point contribution, the stochastic corrections to the tree-level bispectrum are necessary.  The terms in Eq.~\ref{eqn:delta_short} give rise to the contributions
\begin{equation}
    B \supset \left[(B_{\rm shot} b_1 + 2Nf\mu^2)Z_1(\bm{k_1}) P_L(k_1) + {\rm cycl.} \right] + B_0
\end{equation}
where the contributions proportional to $B_{\rm shot}$, $N$, and $B_0$ arise from $\expval{ (\delta) (d_1 \epsilon)(d_2 b_1 \epsilon \delta)}=d_1 d_2 b_1\expval{\epsilon^2}\expval{\delta^2}$, $\expval{ (\delta) (d_1 \epsilon)(d_1 f \mu^2\epsilon \theta)}=d_1^2 f\mu^2 \expval{\epsilon^2}\expval{\delta^2}$ and $\expval{(d_1 \epsilon)^3}$. Here, $N$ is the shot-noise term of the power spectrum and $B_{\rm shot}$, $B_0$ are new degrees of freedom. These additional terms add the following contributions to Eqn.~(\ref{M_B})
\begin{align}
\begin{split}
    M_{13}^B &\supset M_{13,\mathrm{s.t.}}^B \\ &= 2C_{\delta_M}(k) \int_{\bm{p}} C_{\delta_M^2}(p,|\bm{k}-\bm{p}|) \\ &\times [(B_{\rm shot} b_1+ 2N f\mu_1^2)Z_1(\bm{k}) P_L(k) +(B_{\rm shot} b_1+ 2N f\mu_p^2)Z_1(\bm{p}) P_L(p)  + B_0] \\
    &= C_{\delta_M}(k) Z_1(\bm{k}) P_L(k) (B_{\rm shot} b_1+2Nf\mu^2) \int_{\bm{p}} [
C_2 W_R(p)W_R(|\bm{k}-\bm{p}|)+C_1W_R(p) ] \\
    &+  C_{\delta_M}(k)  \int_{\bm{p}}[
2C_2 W_R(p)W_R(|\bm{k}-\bm{p}|)+C_1(W_R(p)+W_R(|\bm{k}-\bm{p}|)) ] \\ &\times (B_{\rm shot} b_1+2Nf\mu_p^2) Z_1(\bm{p}) P_L(p) \\
    &+  C_{\delta_M}(k) B_0 \int_{\bm{p}}[2C_2 W_R(p)W_R(|\bm{k}-\bm{p}|)+2C_1W_R(p) ] \\
    M_{22}^B &\supset M_{22,\mathrm{s.t.}}^B =2M_{13,\mathrm{s.t.}}^B
\end{split}
\label{M_B_correction}
\end{align}
where the $M_{22}^B$ contribution is identical (up to a factor of two) to that of $M_{13}^B$ after a change of variables. 

As the four-point contributions are composed of two `disconnected' linear power spectra, to tree-level the treatment of stochasticity is to simply include the shot noise term to each contribution. Thus the expressions in Eqn.~(\ref{M_C}) transform to
\begin{align}
\begin{split}
    M_{13}^C &= 3 C_{\delta_M}(k) (Z_1(\bm{k})^2 P_L(k)  + N)\int_{\bm{p}} C_{\delta_M^3}(p,p,k)  (Z_1(\bm{p})^2 P_L(k) + N)\\ 
    M_{22}^C &= 2 C_{\delta_M^2}(p,|\bm{k}-\bm{p}|) \int_{\bm{p}} C_{\delta_M^3}(p,p,k) (Z_1(\bm{p})^2 P_L(k)+  N)(Z_1(\bm{k}-\bm{p})^2 P_L(k) + N)     
\end{split}
\label{M_C_correction}
\end{align}
where $N$ is the shot noise term in the power spectrum. 

The one-loop bispectrum corrections modify Eqn.~(\ref{M_B}) such that, e.g.
\begin{align}
    M_{13}^B &= 2C_{\delta_M}(k) \int_{\bm{p}} C_{\delta_M^2}(p,|\bm{k}-\bm{p}|) [Z_1^{\rm FoG}(\bm{k})Z_1^{\rm FoG}(\bm{p})Z_2(\bm{k},-\bm{p})P_L(k)P_L(p) \\
    &+ (B_{\rm shot} b_1+ 2N f\mu^2)Z_1^{\rm FoG}(\bm{k}) P_L(k) +(B_{\rm shot} b_1+ 2N f\mu_p^2)Z_1^{\rm FoG}(\bm{p}) P_L(p)  + B_0] 
\end{align}
with a similar expression for $M_{22}^B$.  These new terms yield the contributions in Eq.~(\ref{eqn:1-loop-stochastic}).

In practice, we find that a simple ansatz of the form 
\begin{equation}
    \bar{m}^2M(k,\mu) \supset C_{\delta_M}(k) k^2 (N_{m,2}^{(0)} + \mu^2 N_{m,2}^{(2)} )
    \label{ansatz_appendix}
\end{equation}
with $N_{m,2}^{(0)}$ and $N_{m,2}^{(2)}$ both new degrees of freedom, retains the key features of the above array of corrections and is sufficient to demonstrate agreement between analytical results and simulations. In total, the stochastic corrections to the marked power spectrum are the tree-level corrections in Eqns.~\ref{M_A_correction_appendix}, \ref{M_B_correction}, and \ref{M_C_correction}, and the ansatz of Eqn.~\ref{ansatz_appendix}.

\bibliographystyle{JHEP}
\bibliography{main}

\providecommand{\href}[2]{#2}\begingroup\raggedright\begin{thebibliography}{10}

\bibitem{Bernardeau02}
F.~{Bernardeau}, S.~{Colombi}, E.~{Gazta{\~n}aga}, and R.~{Scoccimarro}, {\it {Large-scale structure of the Universe and cosmological perturbation theory}},  {\em \physrep} {\bf 367} (Sept., 2002) 1--248, [\href{http://xxx.lanl.gov/abs/astro-ph/0112551}{{\tt astro-ph/0112551}}].

\bibitem{Ferraro22}
S.~{Ferraro}, N.~{Sailer}, A.~{Slosar}, and M.~{White}, {\it {Snowmass2021 Cosmic Frontier White Paper: Cosmology and Fundamental Physics from the three-dimensional Large Scale Structure}},  {\em arXiv e-prints} (Mar., 2022) arXiv:2203.07506, [\href{http://arxiv.org/abs/2203.07506}{{\tt arXiv:2203.07506}}].

\bibitem{Ivanov22b}
M.~M. {Ivanov}, {\it {Effective Field Theory for Large Scale Structure}},  {\em arXiv e-prints} (Dec., 2022) arXiv:2212.08488, [\href{http://arxiv.org/abs/2212.08488}{{\tt arXiv:2212.08488}}].

\bibitem{Baumann22}
D.~Baumann, {\em Cosmology}.
\newblock Cambridge University Press, 2022.

\bibitem{Alam21}
S.~{Alam}, M.~{Aubert}, S.~{Avila}, et~al., {\it {Completed SDSS-IV extended Baryon Oscillation Spectroscopic Survey: Cosmological implications from two decades of spectroscopic surveys at the Apache Point Observatory}},  {\em \prd} {\bf 103} (Apr., 2021) 083533, [\href{http://arxiv.org/abs/2007.08991}{{\tt arXiv:2007.08991}}].

\bibitem{DESI24-VI}
{DESI Collaboration}, A.~G. {Adame}, J.~{Aguilar}, et~al., {\it {DESI 2024 VI: Cosmological Constraints from the Measurements of Baryon Acoustic Oscillations}},  {\em arXiv e-prints} (Apr., 2024) arXiv:2404.03002, [\href{http://arxiv.org/abs/2404.03002}{{\tt arXiv:2404.03002}}].

\bibitem{Peebles75}
P.~J.~E. {Peebles} and E.~J. {Groth}, {\it {Statistical analysis of catalogs of extragalactic objects. V. Three-point correlation function for the galaxy distribution in the Zwicky catalog.}},  {\em \apj} {\bf 196} (Feb., 1975) 1--11.

\bibitem{Fry82}
J.~N. {Fry} and M.~{Seldner}, {\it {Transform analysis of the high-resolution Shane-Wirtanen Catalog - The power spectrum and the bispectrum}},  {\em \apj} {\bf 259} (Aug., 1982) 474--481.

\bibitem{Fry78}
J.~N. {Fry} and P.~J.~E. {Peebles}, {\it {Statistical analysis of catalogs of extragalactic objects. IX. The four-point galaxy correlation function.}},  {\em \apj} {\bf 221} (Apr., 1978) 19--33.

\bibitem{Hu01}
W.~{Hu}, {\it {Angular trispectrum of the cosmic microwave background}},  {\em \prd} {\bf 64} (Oct., 2001) 083005, [\href{http://xxx.lanl.gov/abs/astro-ph/0105117}{{\tt astro-ph/0105117}}].

\bibitem{Gil-Marin16}
H.~{Gil-Mar{\'\i}n}, W.~J. {Percival}, J.~R. {Brownstein}, et~al., {\it {The clustering of galaxies in the SDSS-III Baryon Oscillation Spectroscopic Survey: RSD measurement from the LOS-dependent power spectrum of DR12 BOSS galaxies}},  {\em \mnras} {\bf 460} (Aug., 2016) 4188--4209, [\href{http://arxiv.org/abs/1509.06386}{{\tt arXiv:1509.06386}}].

\bibitem{Philcox22b}
O.~H.~E. {Philcox}, M.~M. {Ivanov}, G.~{Cabass}, et~al., {\it {Cosmology with the redshift-space galaxy bispectrum monopole at one-loop order}},  {\em \prd} {\bf 106} (Aug., 2022) 043530, [\href{http://arxiv.org/abs/2206.02800}{{\tt arXiv:2206.02800}}].

\bibitem{Ivanov23}
M.~M. {Ivanov}, O.~H.~E. {Philcox}, G.~{Cabass}, et~al., {\it {Cosmology with the galaxy bispectrum multipoles: Optimal estimation and application to BOSS data}},  {\em \prd} {\bf 107} (Apr., 2023) 083515, [\href{http://arxiv.org/abs/2302.04414}{{\tt arXiv:2302.04414}}].

\bibitem{DAmico24}
G.~{D'Amico}, Y.~{Donath}, M.~{Lewandowski}, et~al., {\it {The one-loop bispectrum of galaxies in redshift space from the Effective Field Theory of Large-Scale Structure}},  {\em \jcap} {\bf 2024} (July, 2024) 041, [\href{http://arxiv.org/abs/2211.17130}{{\tt arXiv:2211.17130}}].

\bibitem{Bernardeau97}
F.~{Bernardeau}, L.~{van Waerbeke}, and Y.~{Mellier}, {\it {Weak lensing statistics as a probe of \{OMEGA\} and power spectrum.}},  {\em \aap} {\bf 322} (June, 1997) 1--18, [\href{http://xxx.lanl.gov/abs/astro-ph/9609122}{{\tt astro-ph/9609122}}].

\bibitem{Pires12}
S.~{Pires}, A.~{Leonard}, and J.-L. {Starck}, {\it {Cosmological constraints from the capture of non-Gaussianity in weak lensing data}},  {\em \mnras} {\bf 423} (June, 2012) 983--992, [\href{http://arxiv.org/abs/1203.2877}{{\tt arXiv:1203.2877}}].

\bibitem{Hahn21}
C.~{Hahn} and F.~{Villaescusa-Navarro}, {\it {Constraining M$_{{\ensuremath{\nu}}}$ with the bispectrum. Part II. The information content of the galaxy bispectrum monopole}},  {\em \jcap} {\bf 2021} (Apr., 2021) 029, [\href{http://arxiv.org/abs/2012.02200}{{\tt arXiv:2012.02200}}].

\bibitem{Philcox22c}
O.~H.~E. Philcox, {\it Probing parity violation with the four-point correlation function of boss galaxies},  {\em Phys. Rev. D} {\bf 106} (Sep, 2022) 063501.

\bibitem{Giri23}
U.~{Giri}, M.~{M{\"u}nchmeyer}, and K.~M. {Smith}, {\it {Constraining $f_{NL}$ using the Large-Scale Modulation of Small-Scale Statistics}},  {\em arXiv e-prints} (May, 2023) arXiv:2305.03070, [\href{http://arxiv.org/abs/2305.03070}{{\tt arXiv:2305.03070}}].

\bibitem{Choustikov23}
N.~Choustikov, Z.~Vlah, and A.~Challinor, {\it Optimizing the evolution of perturbations in the $\mathrm{\ensuremath{\Lambda}}\mathrm{CDM}$ universe},  {\em Phys. Rev. D} {\bf 108} (Jul, 2023) 023529.

\bibitem{Hou23}
J.~{Hou}, Z.~{Slepian}, and R.~N. {Cahn}, {\it {Measurement of parity-odd modes in the large-scale 4-point correlation function of Sloan Digital Sky Survey Baryon Oscillation Spectroscopic Survey twelfth data release CMASS and LOWZ galaxies}},  {\em \mnras} {\bf 522} (May, 2023) 5701--5739, [\href{http://arxiv.org/abs/2206.03625}{{\tt arXiv:2206.03625}}].

\bibitem{Philcox24}
O.~H.~E. {Philcox} and T.~{Fl{\"o}ss}, {\it {PolyBin3D: A Suite of Optimal and Efficient Power Spectrum and Bispectrum Estimators for Large-Scale Structure}},  {\em arXiv e-prints} (Apr., 2024) arXiv:2404.07249, [\href{http://arxiv.org/abs/2404.07249}{{\tt arXiv:2404.07249}}].

\bibitem{Harscouet24}
L.~{Harscouet}, J.~A. {Cowell}, J.~{Ereza}, et~al., {\it {Fast Projected Bispectra: the filter-square approach}},  {\em arXiv e-prints} (Sept., 2024) arXiv:2409.07980, [\href{http://arxiv.org/abs/2409.07980}{{\tt arXiv:2409.07980}}].

\bibitem{Simpson11}
F.~{Simpson}, J.~B. {James}, A.~F. {Heavens}, and C.~{Heymans}, {\it {Clipping the Cosmos: The Bias and Bispectrum of Large Scale Structure}},  {\em Phys.Rev.Lett.} {\bf 107} (Dec., 2011) 271301, [\href{http://arxiv.org/abs/1107.5169}{{\tt arXiv:1107.5169}}].

\bibitem{Gruen16}
D.~{Gruen}, O.~{Friedrich}, A.~{Amara}, et~al., {\it {Weak lensing by galaxy troughs in DES Science Verification data}},  {\em \mnras} {\bf 455} (Jan., 2016) 3367--3380, [\href{http://arxiv.org/abs/1507.05090}{{\tt arXiv:1507.05090}}].

\bibitem{Friedrich18}
{\bf DES Collaboration} Collaboration, O.~Friedrich, D.~Gruen, J.~DeRose, et~al., {\it Density split statistics: Joint model of counts and lensing in cells},  {\em Phys. Rev. D} {\bf 98} (Jul, 2018) 023508.

\bibitem{Paillas21}
E.~{Paillas}, Y.-C. {Cai}, N.~{Padilla}, and A.~G. {S{\'a}nchez}, {\it {Redshift-space distortions with split densities}},  {\em \mnras} {\bf 505} (Aug., 2021) 5731--5752, [\href{http://arxiv.org/abs/2101.09854}{{\tt arXiv:2101.09854}}].

\bibitem{Paillas24}
E.~{Paillas}, C.~{Cuesta-Lazaro}, W.~J. {Percival}, et~al., {\it {Cosmological constraints from density-split clustering in the BOSS CMASS galaxy sample}},  {\em \mnras} {\bf 531} (June, 2024) 898--918, [\href{http://arxiv.org/abs/2309.16541}{{\tt arXiv:2309.16541}}].

\bibitem{Banerjee21}
A.~{Banerjee} and T.~{Abel}, {\it {Nearest neighbour distributions: New statistical measures for cosmological clustering}},  {\em \mnras} {\bf 500} (Jan., 2021) 5479--5499, [\href{http://arxiv.org/abs/2007.13342}{{\tt arXiv:2007.13342}}].

\bibitem{Cheng20}
S.~{Cheng}, Y.-S. {Ting}, B.~{M{\'e}nard}, and J.~{Bruna}, {\it {A new approach to observational cosmology using the scattering transform}},  {\em \mnras} {\bf 499} (Dec., 2020) 5902--5914, [\href{http://arxiv.org/abs/2006.08561}{{\tt arXiv:2006.08561}}].

\bibitem{Valogiannis22}
G.~Valogiannis and C.~Dvorkin, {\it Towards an optimal estimation of cosmological parameters with the wavelet scattering transform},  {\em Phys. Rev. D} {\bf 105} (May, 2022) 103534.

\bibitem{Eickenberg22}
M.~{Eickenberg}, E.~{Allys}, A.~{Moradinezhad Dizgah}, et~al., {\it {Wavelet Moments for Cosmological Parameter Estimation}},  {\em arXiv e-prints} (Apr., 2022) arXiv:2204.07646, [\href{http://arxiv.org/abs/2204.07646}{{\tt arXiv:2204.07646}}].

\bibitem{Cheng24}
S.~{Cheng}, G.~A. {Marques}, D.~{Grand{\'o}n}, et~al., {\it {Cosmological constraints from weak lensing scattering transform using HSC Y1 data}},  {\em arXiv e-prints} (Apr., 2024) arXiv:2404.16085, [\href{http://arxiv.org/abs/2404.16085}{{\tt arXiv:2404.16085}}].

\bibitem{Rubira21}
H.~{Rubira} and R.~{Voivodic}, {\it {The effective field theory and perturbative analysis for log-density fields}},  {\em \jcap} {\bf 2021} (Mar., 2021) 070, [\href{http://arxiv.org/abs/2011.12280}{{\tt arXiv:2011.12280}}].

\bibitem{Krause24}
{Beyond-2pt Collaboration}, {:}, E.~{Krause}, et~al., {\it {A Parameter-Masked Mock Data Challenge for Beyond-Two-Point Galaxy Clustering Statistics}},  {\em arXiv e-prints} (May, 2024) arXiv:2405.02252, [\href{http://arxiv.org/abs/2405.02252}{{\tt arXiv:2405.02252}}].

\bibitem{Schmittfull15a}
M.~{Schmittfull}, T.~{Baldauf}, and U.~{Seljak}, {\it {Near optimal bispectrum estimators for large-scale structure}},  {\em \prd} {\bf 91} (Feb., 2015) 043530, [\href{http://arxiv.org/abs/1411.6595}{{\tt arXiv:1411.6595}}].

\bibitem{Chen24}
S.-F. {Chen}, P.~{Chakraborty}, and C.~{Dvorkin}, {\it {Analysis of BOSS galaxy data with weighted skew-spectra}},  {\em \jcap} {\bf 2024} (May, 2024) 011, [\href{http://arxiv.org/abs/2401.13036}{{\tt arXiv:2401.13036}}].

\bibitem{Hou24}
J.~{Hou}, A.~M. {Dizgah}, C.~{Hahn}, et~al., {\it {Cosmological constraints from the redshift-space galaxy skew spectra}},  {\em \prd} {\bf 109} (May, 2024) 103528, [\href{http://arxiv.org/abs/2401.15074}{{\tt arXiv:2401.15074}}].

\bibitem{Stoyan84}
D.~Stoyan, {\it On correlations of marked point processes},  {\em Mathematische Nachrichten} {\bf 116} (1984), no.~1 197--207, [\href{http://xxx.lanl.gov/abs/https://onlinelibrary.wiley.com/doi/pdf/10.1002/mana.19841160115}{{\tt https://onlinelibrary.wiley.com/doi/pdf/10.1002/mana.19841160115}}].

\bibitem{Beisbart02}
C.~{Beisbart}, M.~{Kerscher}, and K.~{Mecke}, {\it {Mark Correlations: Relating Physical Properties to Spatial Distributions}},  in {\em Morphology of Condensed Matter} (K.~{Mecke} and D.~{Stoyan}, eds.), vol.~600, pp.~358--390.
\newblock 2002.

\bibitem{Skibba06}
R.~{Skibba}, R.~K. {Sheth}, A.~J. {Connolly}, and R.~{Scranton}, {\it {The luminosity-weighted or `marked' correlation function}},  {\em \mnras} {\bf 369} (June, 2006) 68--76, [\href{http://xxx.lanl.gov/abs/astro-ph/0512463}{{\tt astro-ph/0512463}}].

\bibitem{White09b}
M.~{White} and N.~{Padmanabhan}, {\it {Breaking halo occupation degeneracies with marked statistics}},  {\em \mnras} {\bf 395} (June, 2009) 2381--2384, [\href{http://arxiv.org/abs/0812.4288}{{\tt arXiv:0812.4288}}].

\bibitem{White16}
M.~White, {\it A marked correlation function for constraining modified gravity models},  {\em Journal of Cosmology and Astroparticle Physics} {\bf 2016} (Nov., 2016) 057–057.

\bibitem{Armijo18}
J.~{Armijo}, Y.-C. {Cai}, N.~{Padilla}, et~al., {\it {Testing modified gravity using a marked correlation function}},  {\em \mnras} {\bf 478} (Aug., 2018) 3627--3632, [\href{http://arxiv.org/abs/1801.08975}{{\tt arXiv:1801.08975}}].

\bibitem{Hernandez-Aguayo18}
C.~{Hern{\'a}ndez-Aguayo}, C.~M. {Baugh}, and B.~{Li}, {\it {Marked clustering statistics in f(R) gravity cosmologies}},  {\em \mnras} {\bf 479} (Oct., 2018) 4824--4835, [\href{http://arxiv.org/abs/1801.08880}{{\tt arXiv:1801.08880}}].

\bibitem{Satpathy19}
S.~{Satpathy}, R.~{A C Croft}, S.~{Ho}, and B.~{Li}, {\it {Measurement of marked correlation functions in SDSS-III Baryon Oscillation Spectroscopic Survey using LOWZ galaxies in Data Release 12}},  {\em \mnras} {\bf 484} (Apr., 2019) 2148--2165, [\href{http://arxiv.org/abs/1901.01447}{{\tt arXiv:1901.01447}}].

\bibitem{Armijo24}
J.~{Armijo}, C.~M. {Baugh}, P.~{Norberg}, and N.~D. {Padilla}, {\it {A new test of gravity - I. Introduction to the method}},  {\em \mnras} {\bf 529} (Apr., 2024) 2866--2876, [\href{http://arxiv.org/abs/2304.06218}{{\tt arXiv:2304.06218}}].

\bibitem{Karcher24}
M.~{K{\"a}rcher}, J.~{Bel}, and S.~{de la Torre}, {\it {Towards an optimal marked correlation function analysis for the detection of modified gravity}},  {\em arXiv e-prints} (June, 2024) arXiv:2406.02504, [\href{http://arxiv.org/abs/2406.02504}{{\tt arXiv:2406.02504}}].

\bibitem{Philcox20}
O.~H. Philcox, E.~Massara, and D.~N. Spergel, {\it What does the marked power spectrum measure? insights from perturbation theory},  {\em Physical Review D} {\bf 102} (Aug., 2020).

\bibitem{Philcox21}
O.~H. Philcox, A.~Aviles, and E.~Massara, {\it Modeling the marked spectrum of matter and biased tracers in real- and redshift-space},  {\em Journal of Cosmology and Astroparticle Physics} {\bf 2021} (Mar., 2021) 038.

\bibitem{Cowell24}
J.~A. {Cowell}, D.~{Alonso}, and J.~{Liu}, {\it {Hitting the mark: Optimising Marked Power Spectra for Cosmology}},  {\em arXiv e-prints} (Sept., 2024) arXiv:2409.05695, [\href{http://arxiv.org/abs/2409.05695}{{\tt arXiv:2409.05695}}].

\bibitem{Massara21}
E.~Massara, F.~Villaescusa-Navarro, S.~Ho, et~al., {\it Using the marked power spectrum to detect the signature of neutrinos in large-scale structure},  {\em Physical Review Letters} {\bf 126} (Jan., 2021).

\bibitem{Maus24b}
M.~{Maus}, S.~{Chen}, M.~{White}, et~al., {\it {An analysis of parameter compression and full-modeling techniques with Velocileptors for DESI 2024 and beyond}},  {\em arXiv e-prints} (Apr., 2024) arXiv:2404.07312, [\href{http://arxiv.org/abs/2404.07312}{{\tt arXiv:2404.07312}}].

\bibitem{Ivanov19}
M.~M. {Ivanov}, M.~{Simonovi{\'c}}, and M.~{Zaldarriaga}, {\it {Cosmological Parameters from the BOSS Galaxy Power Spectrum}},  {\em arXiv e-prints} (Sep, 2019) arXiv:1909.05277, [\href{http://arxiv.org/abs/1909.05277}{{\tt arXiv:1909.05277}}].

\bibitem{McDonald00}
P.~{McDonald}, J.~{Miralda-Escud{\'e}}, M.~{Rauch}, et~al., {\it {The Observed Probability Distribution Function, Power Spectrum, and Correlation Function of the Transmitted Flux in the Ly{\ensuremath{\alpha}} Forest}},  {\em \apj} {\bf 543} (Nov., 2000) 1--23, [\href{http://xxx.lanl.gov/abs/astro-ph/9911196}{{\tt astro-ph/9911196}}].

\bibitem{Mcdonald03}
P.~{McDonald}, {\it {Toward a Measurement of the Cosmological Geometry at z \raisebox{-0.5ex}\textasciitilde 2: Predicting Ly{\ensuremath{\alpha}} Forest Correlation in Three Dimensions and the Potential of Future Data Sets}},  {\em \apj} {\bf 585} (Mar., 2003) 34--51, [\href{http://xxx.lanl.gov/abs/astro-ph/0108064}{{\tt astro-ph/0108064}}].

\bibitem{Seljak12}
U.~{Seljak}, {\it {Bias, redshift space distortions and primordial nongaussianity of nonlinear transformations: application to Ly-{\ensuremath{\alpha}} forest}},  {\em \jcap} {\bf 2012} (Mar., 2012) 004, [\href{http://arxiv.org/abs/1201.0594}{{\tt arXiv:1201.0594}}].

\bibitem{Cieplak16}
A.~M. {Cieplak} and A.~{Slosar}, {\it {Towards physics responsible for large-scale Lyman-{\ensuremath{\alpha}} forest bias parameters}},  {\em \jcap} {\bf 2016} (Mar., 2016) 016, [\href{http://arxiv.org/abs/1509.07875}{{\tt arXiv:1509.07875}}].

\bibitem{Chen21}
S.-F. {Chen}, Z.~{Vlah}, and M.~{White}, {\it {The Ly{\ensuremath{\alpha}} forest flux correlation function: a perturbation theory perspective}},  {\em \jcap} {\bf 2021} (May, 2021) 053, [\href{http://arxiv.org/abs/2103.13498}{{\tt arXiv:2103.13498}}].

\bibitem{Ivanov24}
M.~M. {Ivanov}, {\it {Lyman alpha forest power spectrum in effective field theory}},  {\em \prd} {\bf 109} (Jan., 2024) 023507, [\href{http://arxiv.org/abs/2309.10133}{{\tt arXiv:2309.10133}}].

\bibitem{DESI24-IV}
{DESI Collaboration}, A.~G. {Adame}, J.~{Aguilar}, et~al., {\it {DESI 2024 IV: Baryon Acoustic Oscillations from the Lyman Alpha Forest}},  {\em arXiv e-prints} (Apr., 2024) arXiv:2404.03001, [\href{http://arxiv.org/abs/2404.03001}{{\tt arXiv:2404.03001}}].

\bibitem{Chuang17}
C.-H. Chuang, F.-S. Kitaura, Y.~Liang, et~al., {\it Linear redshift space distortions for cosmic voids based on galaxies in redshift space},  {\em Phys. Rev. D} {\bf 95} (Mar, 2017) 063528.

\bibitem{McQuinn11}
M.~{McQuinn} and M.~{White}, {\it {On estimating Ly{\ensuremath{\alpha}} forest correlations between multiple sightlines}},  {\em \mnras} {\bf 415} (Aug., 2011) 2257--2269, [\href{http://arxiv.org/abs/1102.1752}{{\tt arXiv:1102.1752}}].

\bibitem{deBelsunce24}
R.~{de Belsunce}, O.~H.~E. {Philcox}, V.~{Ir{\v{s}}i{\v{c}}}, et~al., {\it {The 3D Lyman-{\ensuremath{\alpha}} forest power spectrum from eBOSS DR16}},  {\em \mnras} {\bf 533} (Sept., 2024) 3756--3770, [\href{http://arxiv.org/abs/2403.08241}{{\tt arXiv:2403.08241}}].

\bibitem{Rogers18}
K.~K. {Rogers}, S.~{Bird}, H.~V. {Peiris}, et~al., {\it {Correlations in the three-dimensional Lyman-alpha forest contaminated by high column density absorbers}},  {\em \mnras} {\bf 476} (May, 2018) 3716--3728, [\href{http://arxiv.org/abs/1711.06275}{{\tt arXiv:1711.06275}}].

\bibitem{Slosar11}
A.~{Slosar}, A.~{Font-Ribera}, M.~M. {Pieri}, et~al., {\it {The Lyman-{\ensuremath{\alpha}} forest in three dimensions: measurements of large scale flux correlations from BOSS 1st-year data}},  {\em \jcap} {\bf 2011} (Sept., 2011) 001, [\href{http://arxiv.org/abs/1104.5244}{{\tt arXiv:1104.5244}}].

\bibitem{Mergulhao23}
T.~Mergulhão, H.~Rubira, and R.~Voivodic, {\it The effective field theory of large-scale structure and multi-tracer ii: redshift space and realistic tracers},  2023.

\bibitem{Ebina24}
H.~{Ebina} and M.~{White}, {\it {Cosmology before noon with multiple galaxy populations}},  {\em \jcap} {\bf 2024} (June, 2024) 052, [\href{http://arxiv.org/abs/2401.13166}{{\tt arXiv:2401.13166}}].

\bibitem{Maksimova21}
N.~A. Maksimova, L.~H. Garrison, D.~J. Eisenstein, et~al., {\it {AbacusSummit: a massive set of high-accuracy, high-resolution N-body simulations}},  {\em Monthly Notices of the Royal Astronomical Society} {\bf 508} (09, 2021) 4017--4037, [\href{http://xxx.lanl.gov/abs/https://academic.oup.com/mnras/article-pdf/508/3/4017/40811763/stab2484.pdf}{{\tt https://academic.oup.com/mnras/article-pdf/508/3/4017/40811763/stab2484.pdf}}].

\bibitem{Garrison18}
L.~H. {Garrison}, D.~J. {Eisenstein}, D.~{Ferrer}, et~al., {\it {The Abacus Cosmos: A Suite of Cosmological N-body Simulations}},  {\em \apjs} {\bf 236} (Jun, 2018) 43, [\href{http://arxiv.org/abs/1712.05768}{{\tt arXiv:1712.05768}}].

\bibitem{Garrison21}
L.~H. {Garrison}, D.~J. {Eisenstein}, D.~{Ferrer}, et~al., {\it {The ABACUS cosmological N-body code}},  {\em \mnras} {\bf 508} (Nov., 2021) 575--596, [\href{http://arxiv.org/abs/2110.11392}{{\tt arXiv:2110.11392}}].

\bibitem{PCP18}
{Planck Collaboration}, N.~{Aghanim}, Y.~{Akrami}, et~al., {\it {Planck 2018 results. VI. Cosmological parameters}},  {\em arXiv e-prints} (July, 2018) arXiv:1807.06209, [\href{http://arxiv.org/abs/1807.06209}{{\tt arXiv:1807.06209}}].

\bibitem{Maus24}
M.~{Maus}, Y.~{Lai}, H.~E. {Noriega}, et~al., {\it {A comparison of effective field theory models of redshift space galaxy power spectra for DESI 2024 and future surveys}},  {\em arXiv e-prints} (Apr., 2024) arXiv:2404.07272, [\href{http://arxiv.org/abs/2404.07272}{{\tt arXiv:2404.07272}}].

\bibitem{abacusHOD}
S.~Yuan, L.~H. Garrison, B.~Hadzhiyska, et~al., {\it {AbacusHOD: a highly efficient extended multitracer HOD framework and its application to BOSS and eBOSS data}},  {\em Monthly Notices of the Royal Astronomical Society} {\bf 510} (11, 2021) 3301--3320, [\href{http://xxx.lanl.gov/abs/https://academic.oup.com/mnras/article-pdf/510/3/3301/42147651/stab3355.pdf}{{\tt https://academic.oup.com/mnras/article-pdf/510/3/3301/42147651/stab3355.pdf}}].

\bibitem{Zheng11}
Z.~{Zheng}, R.~{Cen}, H.~{Trac}, and J.~{Miralda-Escud{\'e}}, {\it {Radiative Transfer Modeling of Ly{\ensuremath{\alpha}} Emitters. II. New Effects on Galaxy Clustering}},  {\em \apj} {\bf 726} (Jan., 2011) 38, [\href{http://arxiv.org/abs/1003.4990}{{\tt arXiv:1003.4990}}].

\bibitem{Marinucci24}
M.~{Marinucci}, G.~{Jung}, M.~{Liguori}, et~al., {\it {The constraining power of the Marked Power Spectrum: an analytical study}},  {\em arXiv e-prints} (Nov., 2024) arXiv:2411.14377, [\href{http://arxiv.org/abs/2411.14377}{{\tt arXiv:2411.14377}}].

\end{thebibliography}\endgroup

\end{document}